\documentclass[usenames,dvipsnames]{aa}
\usepackage{graphicx}
\graphicspath{{./}}
\usepackage[varg]{txfonts}
\usepackage{natbib}
\bibpunct{(}{)}{;}{a}{}{,}
\usepackage{xcolor}
\usepackage{siunitx}
\usepackage[export]{adjustbox}
\usepackage{subcaption}

\newcommand{\diff}{\mathrm{d}}

\usepackage{hyperref}
\hypersetup{colorlinks=true, linkcolor=Maroon,
      citecolor=PineGreen,
      bookmarksnumbered=true, pdfborder={0 0 0},unicode,breaklinks}

\begin{document}

\title{SNe Ia from double detonations: Impact of core-shell mixing on the
carbon ignition mechanism}

\author{Sabrina~Gronow\inst{1,2}\fnmsep\thanks{\email{sabrina.gronow@h-its.org},
Fellow of the International Max Planck Research School for Astronomy and Cosmic
Physics at the Heidelberg University (IMPRS-HD)} \and
Christine~Collins\inst{3} \and
Sebastian~T.~Ohlmann\inst{4,1} \and
R\"udiger~Pakmor\inst{5} \and
Markus~Kromer\inst{1,6} \and
Ivo R. Seitenzahl\inst{7} \and
Stuart~A.~Sim\inst{3} \and
Friedrich~K.~R\"{o}pke\inst{1,6}}

\titlerunning{On core-shell mixing in sub-M$_{\text{Ch}}$ white dwarfs}
\authorrunning{Gronow et al.}

\institute{
Heidelberger Institut f\"{u}r Theoretische Studien,
Schloss-Wolfsbrunnenweg 35, 69118 Heidelberg, Germany
\and
Zentrum f\"ur Astronomie der Universit\"at Heidelberg,
Astronomisches Rechen-Institut, M\"{o}nchhofstr. 12-14, 69120 Heidelberg, Germany
\and
Astrophysics Research Center, School of Mathematics and Physics, Queen's
University Belfast, Belfast BT7 1NN, Northern Ireland, UK
\and
Max Planck Computing and Data Facility, Gie{\ss}enbachstra{\ss}e 2, 85748 Garching, Germany
\and
Max Planck Institute for Astrophysics, Karl-Schwarzschild-Straße 1, 85748, Garching, Germany
\and
Zentrum f\"ur Astronomie der Universit\"at Heidelberg, Institut f\"ur
Theoretische Astrophysik, Philosophenweg 12, 69120 Heidelberg, Germany
\and
School of Science, University of New South Wales, Australian Defence Force
Academy, Canberra, ACT 2600, Australia}

\date{Received August 10, 2019 / Accepted January 23, 2020}

\abstract
{ Sub-Chandrasekhar mass white dwarfs accreting a helium shell on a
carbon-oxygen core are potential progenitors of normal Type Ia supernovae.
This work focuses on the details of the onset of the carbon detonation in the
    double detonation sub-Chandrasekhar model. In order to simulate the
    influence of core-shell mixing on the carbon ignition mechanism, the helium
    shell and its detonation are followed with an increased resolution compared to
the rest of the star treating the propagation of the detonation wave more
accurately. This significantly improves the predictions of the nucleosynthetic
yields from the helium burning.
The simulations were carried out with the \textsc{Arepo} code. A carbon-oxygen
    core with a helium shell was set up in one dimension and mapped to three
    dimensions. We ensured the stability of the
    white dwarf with a relaxation step before the hydrodynamic detonation simulation
started. Synthetic observables were calculated with the radiative transfer code
\textsc{Artis}.
An ignition mechanism of the carbon detonation was observed, which received
    little attention before. In this "scissors mechanism", the impact the helium
    detonation wave has on unburnt material when converging opposite to its
    ignition spot is strong enough to ignite a carbon detonation. This is
    possible in a carbon enriched transition region between the core and shell.
    The detonation mechanism is found to be sensitive to details of the
    core-shell transition and our models illustrate the need to consider
    core-shell mixing taking place during the accretion process. Even though
    the detonation ignition mechanism differs form the converging shock
    mechanism, the differences in the synthetic observables are not significant.
    Though they do not fit observations better than previous simulations, they
    illustrate the need for multi-dimensional simulations. }


\maketitle

\section{Introduction}
\label{sec:introduction}

The progenitor evolution and the conditions that lead to the onset of explosions
of Type Ia supernovae (SNe~Ia) are highly controversial. Recent results
\citep[e.g.,][]{gilfanov2010a,sim2013a} indicate that the majority of events
cannot be easily explained with the long-time favored model of thermonuclear
explosions in Chandrasekhar-mass white dwarfs (WDs)
(\citealp{arnett1969a,reinecke2002d,seitenzahl2013a}, but also see
\citealp{seitenzahl2013b}). A promising alternative is a thermonuclear
carbon-oxygen (CO) detonation in sub-Chandrasekhar mass WDs
\citep[]{shigeyama1992a,sim2010a,shen2018a,wilk2018a}, but the mechanism by
which it is initiated is not fully understood. Violent mergers of two white
dwarfs have been suggested
\citep[e.g.,][]{iben1984a,guillochon2010a,pakmor2010a,pakmor2011b,pakmor2013a}.
As an alternative to that, \citet{nomoto1982b}, \citet{woosley1994a},
\citet{bildsten2007a}, and \citet{kromer2010a} investigated sub-Chandrasekhar
mass white dwarfs in binaries. In this case a white dwarf accretes helium (He) from
a companion and ignition conditions arise. Recent reviews on the different
progenitor systems are \citet{maoz2014a} and \citet{wang2012a}.\\

Here we consider a sub-Chandrasekhar mass white dwarf as a progenitor that
explodes in the double detonation scenario. In this scenario, a carbon-oxygen
WD has accreted a rather massive helium shell from a helium white dwarf. A
detonation is ignited at the base of the shell when critical conditions are
reached through thermal instability. Following the detonation in the shell,
three main scenarios have been suggested so far: First, the He shell detonation
directly triggers a second detonation at the interface between the He shell and
CO core, which is referred to as the edge-lit scenario
\citep[e.g.,][]{livne1990b,sim2012a}. Second, a shock wave that is driven by
the helium shell detonation propagates into the core and converges spherically
thus igniting a second detonation, which is referred to as the converging shock
scenario
\citep[e.g.,][]{livne1990a,livne1991a,livne1995a,fink2007a,fink2010a,moll2013a,shen2014a}.
Third, no secondary core detonation is ignited and a faint .Ia supernova ensues
\citep[e.g.,][]{bildsten2007a, waldman2011a, sim2012a}.

The converging shock scenario corresponds to the classical double detonation
scenario. In the context of this scenario three questions remain open; (1) how
does the helium detonation in the shell form, (2) how is the core detonation
initiated, and (3) what are the yields of the helium detonation and their
effects on the spectra and light curves.

Previous work by \citet{kromer2010a}, \citet{boyle2017a}, and
\citet{botyanszki2018a}, for example, points out that the synthetic spectra of
the sub-Chandrasekhar mass models are too red. Further He is not present in the
observed spectra. Therefore the mass of the He shell must be small.
\citet{townsley2019a} show that a double detonation of a CO white dwarf with a
less massive He shell is possible. Their model is modestly enriched with C
which supports burning to heavier elements during the He detonation. Compared
to their work we now increase the resolution in the He shell and use a
different numerical approach.

Our simulations focus on the last two questions mentioned above. We study the
influence of the burning products of the helium shell on the ejecta composition
and the details of igniting a detonation in the CO core. Radiative transfer
calculations allow the discussion of their impact on the observables. Relating
to question (1), \citet{glasner2018a} investigated whether a helium detonation
can ignite in the He shell. \citet{roepke2007a} and \citet{seitenzahl2009b}
studied conditions for carbon detonation ignition and the processes leading to
it were simulated by \cite{fink2007a, fink2010a} and \cite{ moll2013a},
addressing question (2). The effect of the He shell ejecta on the observables
was discussed by \citet{kromer2010a} and \cite{townsley2012a,townsley2019a},
corresponding to question (3).

Previous work was mostly carried out in one or two spatial dimensions (1D and
2D). \citet{moll2013a} perform three-dimensional (3D) simulations of one
quarter of the white dwarf assuming two synchronous spherical detonators and
using mirror symmetry. \citet{garcia2018b} carry out 3D simulations of rigidly
rotating sub-Chandrasekhar mass white dwarfs with a smoothed particle
hydrodynamics code. We follow up on such studies and present grid-based
hydrodynamic simulations comprising a whole (non-rotating) white dwarf to check
whether a detonation in a helium shell can trigger a second detonation in the
core leading to a complete incineration of the white dwarf and a supernova
explosion. The use of the moving mesh code \textsc{Arepo} \citep{springel2010a}
allows a better resolution of the helium shell and a more accurate simulation
of the propagation of the detonation front in the helium shell compared to
previous simulations by others \citep[e.g.,][]{fink2007a,moll2013a}. In our
simulations we observe a mechanism for igniting the secondary core detonation
that previously received little attention. The convergence of the helium
detonation wave on the far side of the ignition spot causes a second detonation
at the edge of the CO core. This detonation propagates through the whole core
disrupting the white dwarf. \citet{livne1995a} and \citet{garcia1999a} mention
a delayed edge-lit detonation with a second detonation forming at the antipode
of the He ignition point, and \citet{forcada2007a} presents a simulation
showing this effect. A detailed discussion of the progenitor and explosion
mechanism, however, is missing in these publications. The same detonation
ignition mechanism is described in \citet{garcia2018b} for a rotating white
dwarf. Following up on their work we explain the ignition mechanism in detail
and discuss its effect on observables in the analyzes of multi-dimensional
radiative transfer calculations.

The methods are described in Section~\ref{sec:methods} and details on the
models are presented in Section~\ref{sec:models}. Section~\ref{sec:results}
discusses the results from the hydrodynamic simulations and their significance
in the framework of previous work. Radiative transfer calculations are
presented in Section~\ref{sec:RT}. We conclude in Section~\ref{sec:summary}.
The simulation data of all our models will be made available on the Heidelberg
Supernova Model Archive \citep[HESMA,][]{kromer2017a}.

\section{Methods}
\label{sec:methods}

\subsection{Hydrodynamics}
\label{sec:hydro}
Three dimensional simulations are carried out using the moving mesh code
\textsc{Arepo} \citep{springel2010a}. The code is based on a Voronoi
tessellation of space with mesh generating points moving along with the
hydrodynamic flow. This leads to a nearly Lagrangian scheme.
A second-order finite-volume method is employed to solve the Euler-Poisson
equations of hydrodynamics with tree-based self-gravity as a source term. The
Godunov method is used as described in \citet{springel2010a} with the improved
scheme of \citet{pakmor2016a}. For modeling reactive flows, a source term is
added to the energy equation and balance equations for nuclear species are
followed. To this end, the nuclear network solver of \citet{pakmor2012b} is
coupled to the hydrodynamic solver of the \textsc{Arepo} code
\citep{pakmor2013a}. If not stated otherwise, we employ a nuclear network
consisting of 33 species comprising n, p, $^4$He, $^{12}$C, $^{13}$N, $^{16}$O,
$^{20}$Ne, $^{22}$Na, $^{23}$Na, $^{24}$Mg, $^{25}$Mg, $^{26}$Mg, $^{27}$Al,
$^{28}$Si, $^{29}$Si, $^{30}$Si, $^{31}$P, $^{32}$S, $^{36}$Ar, $^{40}$Ca,
$^{44}$Ti, $^{45}$Ti, $^{46}$Ti, $^{47}$V, $^{48}$Cr, $^{49}$Cr, $^{50}$Cr,
$^{51}$Mn, $^{52}$Fe, $^{53}$Fe, $^{54}$Fe, $^{55}$Co, and $^{56}$Ni.

Following \citet{fryxell1989a} and Appendix A of \citet{townsley2016a},
    burning is disabled when the conditions
\begin{align}
    \vec{\nabla}\cdot \vec{v} < 0 \: \text{and} \:
    \nabla P \cdot \frac{r_{\text{cell}}}{P_{\text{cell}}} > 0.66
\end{align}
indicate that the corresponding region is located inside the shock. The
Helmholtz equation of state was implemented by \citet{pakmor2013a} based on
\citet{timmes2000a} and closes the system of equations to be solved.

The adaptive mesh refinement capability of \textsc{Arepo} allows us to better
resolve the helium shell and the propagation of the detonation wave within it.
We employ an additional refinement in two different regions: the He shell and
the location of the carbon detonation ignition (see Sec.~\ref{sec:detonation}
for the coordinates). A passive scalar is used to track the location of the He
shell. This is needed as He is not only present in the shell, but also in the
background of the white dwarf.

The mass of a cell is chosen as the refinement criterion. Similar to
\citet{pakmor2013a} an explicit refinement is used for a better mass resolution
and a reference mass for a cell is fixed. If a cell mass exceeds this reference
mass by a factor of two the cell is split.

\subsection{Nucleosynthesis postprocessing}
\label{sec:theory_pp}
As stated above a 33 isotope nuclear network is used to follow the
nucleosynthesis in the hydrodynamics simulation. This gives a good
approximation to the final abundances and the energy release during the
explosion. To determine detailed yields, nucleosynthesis postprocessing
is carried out in a subsequent step based on two million tracer particles that
are placed into the hydrodynamic explosion model. These sample the initial mass
distribution, each representing a mass of about $1\times10^{27}$\,g of material
whose temperature and density evolution is tracked \citep{travaglio2004a}.

The thermodynamic tracer particle trajectories form the basis of the
postprocessing nucleosynthesis calculation with a network involving 384
isotopes \citep{pakmor2012b}. It reaches from neutrons to $^{98}$Mo. The
reaction rates are taken from the REACLIB data base \citep{rauscher2000a} as in
\citet{pakmor2012b}. The hydrodynamic simulation extends up to 100\,s after
ignition. By this time, homologous expansion is reached to a good approximation
while the nuclear burning is already complete after a few seconds. With its two
million tracer particles, the postprocessing step determines the
three-dimensional chemical composition of the ejecta with sufficient accuracy
for subsequent radiative transfer calculations.

\subsection{Radiative transfer}
The time-dependent multi-dimensional Monte Carlo radiative transfer code
\textsc{Artis} (\citealt{sim2007b, kromer2009a}, based on the methods of
\citealt{lucy2002a, lucy2003a, lucy2005a}) is used to derive synthetic
observables for the models. The postprocessing abundances and final ejecta
density are mapped onto a 50$^3$ Cartesian grid using the scheme described by
\citet[][see also \citealt{kromer2010a}]{fink2014a}.  In each radiative
transfer simulation $2.56\times10^7$ energy packets are tracked as they
propagate through the ejecta for 111 logarithmically spaced time steps between
2 and 120 days after the explosion. We use the atomic data set as described by
\cite{gall2012a}, adopt a gray approximation in cells that are optically thick
(cf.  \citealt{kromer2009a}), and assume local thermodynamic equilibrium for
the first ten time steps (times < 3 days after explosion).  Line-of-sight
dependent light curves are calculated as by \citet{kromer2010a}. The escaping
photons are binned into a grid of ten equal solid-angle bins in $\mathrm{\mu =
cos \ \theta}$ where $\theta$ is the angle between the line of sight and the
z-axis of the model. The line-of-sight dependent spectra are calculated using
"virtual-packets" as described by \cite{bulla2015a}.  Where an energy packet
interaction occurs, a virtual-packet is created with frequency and energy equal
to the energy packet at the point of creation. The virtual-packet is propagated
toward a predefined observer direction $n_{\mathrm{obs}}$, and contributes to
the emergent spectrum.  This approach significantly reduces the Monte Carlo
noise in the angle dependent spectra.

\section{Models}
\label{sec:models}

\subsection{Model setup}
\label{sec:setup}
Nine hydrodynamics simulations were carried out following the evolution of a
white dwarf that consists of a carbon-oxygen core and a helium shell. Carbon
and oxygen make up 50\% by mass, respectively, of the core material. In our
initial models, the composition changes in a small transition region at the
edge of the core to pure helium. The details of the models are listed in
Table~\ref{tab:models}. Their total masses of 0.91\,M$_\odot$ (M3a) and
1.05\,M$_\odot$ (M1 and M2) are chosen to be similar to Models 1 and 3
(hereafter FM1 and FM3) in \citet{fink2010a}. Model M2a is our reference model.
Modifications of the base setup are made to study the effect of different
parameters. The different models explore the influence of the mixing of carbon
into the shell (M1a and M2a), the white dwarf mass (M2a and M3a), the
resolution (M2a, M2a\_13, M2a\_21, M2a\_36, and M2a\_79), the nuclear network in
the hydrodynamic calculations (M2a and M2a\_i55), and the He ignition setup (M2a
and b; see Table~\ref{tab:models}).  The reference model M2a is discussed in
Sections \ref{sec:relax} and \ref{sec:detonation} as well as
Sections~\ref{sec:mechanism} and \ref{sec:abundances}. A comparison and
discussion of the hydrodynamic simulations in the context of Models M1a and M2b
to M3a follows in Sections~\ref{sec:mixing} to \ref{sec:comparison}.

The white dwarf was set up in 1D by integrating the equations of hydrostatic
equilibrium. The total mass of the white dwarf ($M_\mathrm{tot}$) and the
transition density ($\rho_\mathrm{S}$) at which the helium shell begins are
initial parameters. The position at which the helium shell begins as well as
its mass depend on these parameters. The temperature was set to be constant
within the core ($T_\mathrm{C}$) while $T_\mathrm{S}$ describes the temperature
at the outer edge of the transition between core and shell beyond which it
declines adiabatically. The helium shell mass was not set explicitly but was
determined by an iteration based on the fixed total mass of the WD and
density at the base of the helium shell, $\rho_\mathrm{S}$, while the central
density $\rho_\mathrm{C}$ is variable.

\begin{figure}[tbp]
\centering
\includegraphics[scale=0.24]{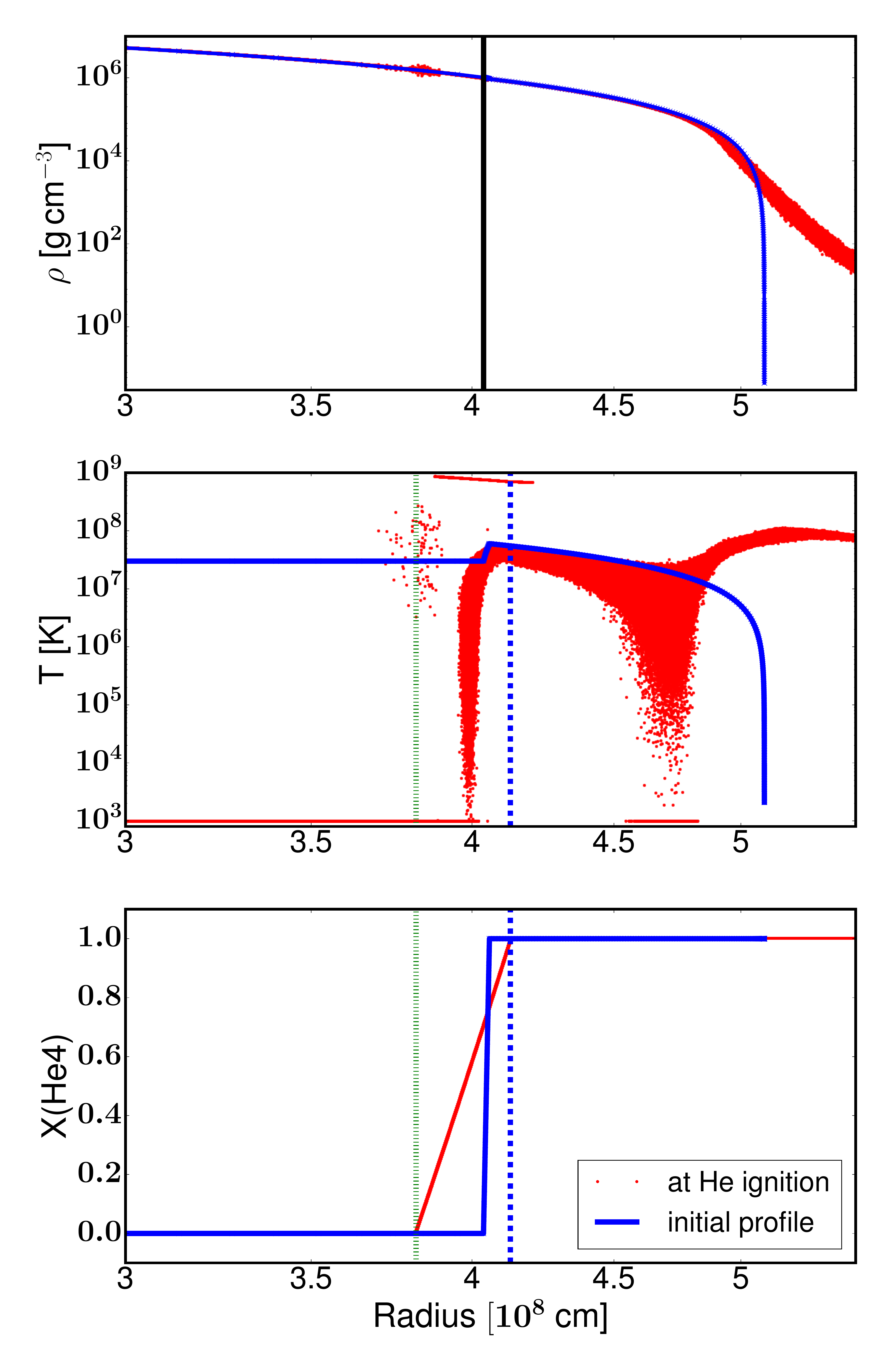}
\caption{Radial profile of the density, temperature, and helium mass fraction in
    an interval of 3~to~$5.5\,~\times\,~10^8$\,cm of
    the initial setup and at helium ignition of Model M2a; the black solid,
    green dotted, and blue dashed lines represent the core-shell transition,
base of the helium shell, and outer edge of the transition region, respectively;
the cells with temperatures higher than $7\times10^8$\,K in the profile at He
ignition represent the detonating cells of the He detonation.}
\label{fig:ini_setup}
\end{figure}

Fig.~\ref{fig:ini_setup} shows the initial setup of Model M2a in the radial
range from 3 to $5.5\times 10^8$\,cm (in blue). The model parameters are listed
in Table~\ref{tab:models}. The jump in the temperature profile (middle row) and
helium mass fraction (bottom row) indicates the location of the change from the
CO core to the helium shell of the WD. The vertical line in the density profile
marks the position of $\rho_S$ corresponding to the other profiles. The core
radius is at about $4.0\times 10^8$\,cm. The transition between core and shell
is set to only consist of 20 cells in the 1D setup following a linear trend in
the temperature and abundances. This transition region extends from
$4.039\times 10^8$\,cm to $4.058\times 10^8$\,cm in radius. Based on the setup
in 1D, the model of a white dwarf with a helium shell is mapped to 3D using the
HEALPix method \citep{gorski2005a} on concentric shells according to
\citet{ohlmann2017a}.

\begin{table*}[htbp]
\caption{Overview of model parameters.}
\label{tab:models}
\centering
\resizebox{\textwidth}{!}{
\begin{tabular}{@{}lrrrrrrrrrr@{}}
\hline\hline
Model & $M_{\mathrm{tot}}$ & $M_{\mathrm{iHeS}}$ & $M_{\mathrm{pHeS}}$ & $T_\mathrm{S}$
      & $T_{\mathrm{C}}$ & $\rho_\mathrm{S}$ & $\rho_\mathrm{C}$ & resolution & \# isotopes & ignition spot \\
      & [M$_\odot$] & [M$_\odot$] & [M$_\odot$] & [$10^7$\,K] & [$10^7$\,K] &
[$10^6$\,$\text{g cm}^{-3}$] & [$10^7$\,$\text{g cm}^{-3}$] &
[$10^{-8}$\,$M_\odot$] & & \\ \hline
M1a& 1.05 & 0.051 & 0.064 & 6 & 3 & 1.2 & 4.8 & 3.33 & 33 & a \\
M2a & 1.05 & 0.051 & 0.073 & 6 & 3 & 1.2 & 4.8 & 3.35 & 33 & a \\
M2b & 1.05 & 0.051 & 0.073 & 6 & 3 & 1.2 & 4.8 & 3.35 & 33 & b \\
M2a\_79 & 1.05 & 0.051 & 0.073 & 6 & 3 & 1.2 & 4.8 & 79.18 & 33 & a \\
M2a\_36 & 1.05 & 0.051 & 0.073 & 6 & 3 & 1.2 & 4.8 & 36.27 & 33 & a \\
M2a\_21 & 1.05 & 0.051 & 0.073 & 6 & 3 & 1.2 & 4.8 & 21.44 & 33 & a \\
M2a\_13 & 1.05 & 0.051 & 0.073 & 6 & 3 & 1.2 & 4.8 & 12.71 & 33 & a \\
M2a\_i55 & 1.05 & 0.051 & 0.073 & 6 & 3 & 1.2 & 4.8 & 3.35 & 55 & a \\
M3a & 0.91 & 0.135 & 0.155 & 6 & 3 & 1.5 & 1.9 & 2.76 & 33 & a \\
    \hline
    \end{tabular}
}
\tablefoot{The total mass of the white dwarf $M_{\mathrm{tot}}$, initial mass
    of the helium shell $M_{\mathrm{iHeS}}$ and post-relaxation
    $M_{\mathrm{pHeS}}$, temperature $T_{\mathrm{S}}$ at the base of the helium
    shell, core temperature $T_{\mathrm{C}}$, transition density to the shell
    $\rho_\mathrm{S}$, and central density $\rho_\mathrm{C}$ are given.
    Models M2 denote models with an additional mixing of carbon into
    the He shell after relaxation. The resolution is given for the volume
    where the detonation wave converges at the antipodes which is the region
    with the highest level of refinement in Model M2a.}
\end{table*}

\subsection{Relaxation}
\label{sec:relax}
After mapping onto the 3D grid of \textsc{Arepo}, a relaxation step following
\citet{ohlmann2017a} was carried out to eliminate spurious velocities
that can be caused by discrepancies between the pressure gradient and gravity.
Earlier work by, for example, \citet{fink2007a} performed the explosion simulations
without prior relaxation of the star; however, when mapping the initial WD
structure onto the unstructured computational mesh of \textsc{Arepo},
relaxation is necessary to obtain a hydrostatic equilibrium. The white dwarf was
relaxed in a hydrodynamic simulation with the nuclear reactions switched off
for ten dynamical time scales, which are defined here by the sound
crossing time,
\begin{equation}
    \tau_\mathrm{dyn} = \int_0^R\frac{\diff r}{v_s}\;,
\end{equation}
with radius $R$ and local sound speed $v_s$, which, because the density is not
constant, varies with radius. In the relaxation process the velocities were
damped until 80\% of the relaxation time has passed. The simulation was
continued for two dynamical timescales without damping to verify stability of
the model. The conditions for stability posed in \citet{ohlmann2017a} are
fulfilled for all models presented in this paper.

During the relaxation process mixing took place between the helium shell and
CO core washing out the interface between them. This mixing was
additionally enhanced after the relaxation for all models except M1a and M3a.
We therefore identify the shell with material in which the helium mass fraction
exceeds $0.01$. Naturally, this shifts the original core and shell masses and
the values for the post-relaxation helium shell masses are included in
Table~\ref{tab:models}.  They differ by about $0.02$\,M$_\odot$ from the
initial setup. $0.01$\,M$_\odot$ of each carbon and oxygen are mixed into the
shell.

The effect of mixing during the relaxation phase is illustrated in
Fig.~\ref{fig:ini_setup}. The red points represent the radial profiles obtained
from the \textsc{Arepo} model at the time of helium ignition after relaxation
while the initial 1D model is shown in blue. The core radius shifts inwards to
about $3.8\times 10^8$\,cm and the transition region between core and
pure shell material increases in size (see bottom panel in
Fig.~\ref{fig:ini_setup}). The radius of the outer end of the transition region
shifts outwards due to the additional artificial mixing. The
temperature profile (middle panel) shows a similar shift corresponding to the
helium mass fraction. However, the transition region does not show an equally
broad increase in the temperature. Some cells in the radial interval
$[3.5\times10^8,3.9\times10^8]$ cm experience an increase in
temperature during the relaxation. This does not influence the detonation as
the values are too low for ignition of a detonation in the material.

\subsection{Detonation}
\label{sec:detonation}
In contrast to \citet{pakmor2013a} who model a white dwarf merger, the first
detonation in the He shell was not ignited dynamically but it was assumed to be
triggered by thermal instability. \citet{glasner2018a} investigate whether
such a detonation in the helium shell can develop naturally following the
accretion of helium onto the white dwarf and find this to be likely from their
simulations. \citet{jacobs2016a} carry out 3D simulations of one eighth
of a WD using a low-Mach number code. They conclude that a localized runaway
can be achieved taking 3D convection into account. Similar to
\citet{livne1997a}, \citet{fink2010a}, and \citet{woosley2011b}, we triggered the
helium detonation artificially at the base of the helium shell. This is
achieved by increasing the specific thermal energy of selected cells to
$5\times 10^{16}$\,erg g$^{-1}$. This value was chosen to ignite
explosive burning without reaching a non-physical high thermal energy.
Comparable values are found by \citet{glasner2018a} in their He detonation
ignition simulations.

Due to the mixing the transition region between core and shell has
widened. The igniting cells were therefore chosen to enclose a volume of radius
$\Delta R$ centered at the outer edge of the transition region. $\Delta R$ was
defined as $0.04$ times the distance of the central detonating cell to the
center of the white dwarf. This value leads to a detonation ignition
in a volume similar to that found by \citet{glasner2018a}. This construction
formally results in a symmetric detonation ignition volume around one point. In
the numerical implementation, however, asymmetries are possible due to the
Voronoi structure of the grid. Averaged global spherical symmetry (neglecting
rotation) allows us to select the detonation point arbitrarily on a sphere
corresponding to the base of the helium shell. In our models, it was placed on
the positive $z$-axis, that is at $x=y=0$.

The cells in which the helium was ignited can be identified in the temperature
profile of Fig.~\ref{fig:ini_setup}. They are located at radii between
$3.88\times10^8$\,cm and $4.11\times10^8$\,cm and have temperatures of at least
$7\times10^8$\,K. The central ignition spot is at a radius of
$4.04\times10^8$\,cm for Model M2a.

\begin{figure}[htbp]
    \centering
    \includegraphics[scale=0.4]{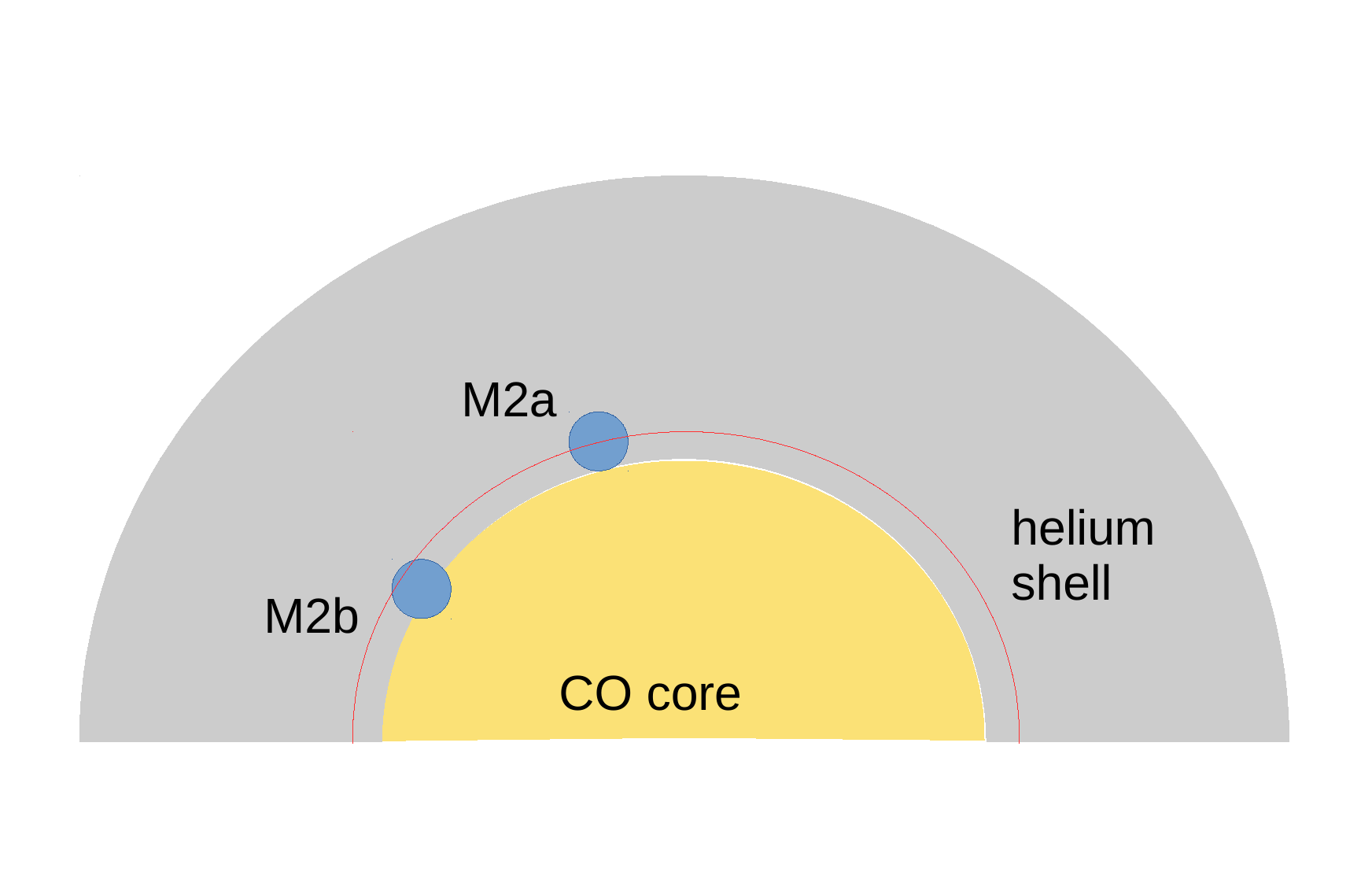}
    \caption{Schematic illustration of different ignition spots for Model
        M2 with the white dwarf CO core in yellow and helium shell in
        gray; symmetric around the base of the shell (M2b) and
        symmetric around the point of peak temperature (M2a).}
    \label{fig:position}
\end{figure}

Different initial positions for the helium detonation were tested for Model M2.
The detonation in the helium shell was ignited uniformly around one point at the
very base of the He shell ($R_\mathrm{C}$, M2b with core radius $R_\mathrm{C}$
after relaxation, green dotted line in Fig.~\ref{fig:ini_setup}), and at the base
of the He shell where the peak in the temperature profile is located at the
outer edge of the transition region (M2a, blue dashed line in
Fig.~\ref{fig:ini_setup}). The radii of the central ignition spots are
$4.04\times10^8$\,cm and $3.77\times10^8$\,cm for Models M2a and b,
respectively. A sketch of the ignition spots is shown in
Fig.~\ref{fig:position}. The red curve represents the outer edge of the
transition region between core and shell.

As described in Section~\ref{sec:hydro} an additional mesh refinement was
imposed in the helium shell to better resolve the propagation of the detonation
wave. A passive scalar was used to follow the shell material and to determine
where the refinement is increased. As we discuss in Sec.~\ref{sec:results},
Model M2a shows a high increase of the density and temperature in the
convergence point of the He detonation wave at the antipode of the helium
ignition which triggers a carbon detonation. A further mesh refinement region
was added for the location of this second, carbon detonation.  This results in
two regions with additional refinement over the base resolution: one in the
helium shell and another around the helium detonation convergence spot opposite
from its ignition. Additional simulations were carried out with different
resolutions to check for the convergence of the helium detonation and the
carbon detonation. These are Models M2a\_79, M2a\_36, M2a\_21, and M2a\_13.
Contrary to simulation M2a and other models no additional refinement is used
for Model M2a\_79.

In \textsc{Arepo}, refinement is enforced by reducing the reference mass in the
respective region. As a standard, a reference mass $M_\text{R}$ of
$2\times10^{27}$\,g is chosen in regions with base resolution. With each level
of refinement the reference mass is decreased. The reference mass is
$4\times10^{26}$\,g and $2\times10^{26}$\,g for Models M2a\_36 and M2a\_21,
respectively, which have an additional refinement in the helium shell. The
refinement around the convergence spot of the helium detonation is imposed in
the region enclosed in $-2\times10^8\,\text{cm} < \text{x} <
2\times10^8\,\text{cm} $, $-2\times10^8\,\text{cm} < \text{y} <
2\times10^8\,\text{cm} $, and $-7\times10^8\,\text{cm} < \text{z} <
-3\times10^8\,\text{cm} $. The limits of this refinement region are chosen
after the first detection of the carbon detonation at the convergence spot.
Models M2a\_13 and M2a have a reference mass of $1.2\times10^{26}$\,g and
$2\times10^{25}$\,g, respectively.

\section{Results from the hydrodynamic explosion simulations}
\label{sec:results}
We summarize the results from the hydrodynamic explosion simulations in this
section focusing on our reference model M2a for most parts. The detonation
ignition mechanism is described in detail followed by an analysis of the final
abundances. A discussion of different parameters such as the resolution,
ignition spot, and white dwarf mass is carried out. A comparison to previous
work concludes this section.

We describe the evolution for our reference model M2a first. After the He
detonation is initiated, a detonation wave propagates through the helium shell
and a shock wave develops and propagates through the core of the white dwarf. The
simulation follows the evolution for 100\,s. In Model M2a a volume consisting
of 4514 cells is set to detonate initially in the helium detonation.\\

\begin{figure*}
  \centering
  \includegraphics[width=0.97\textwidth]{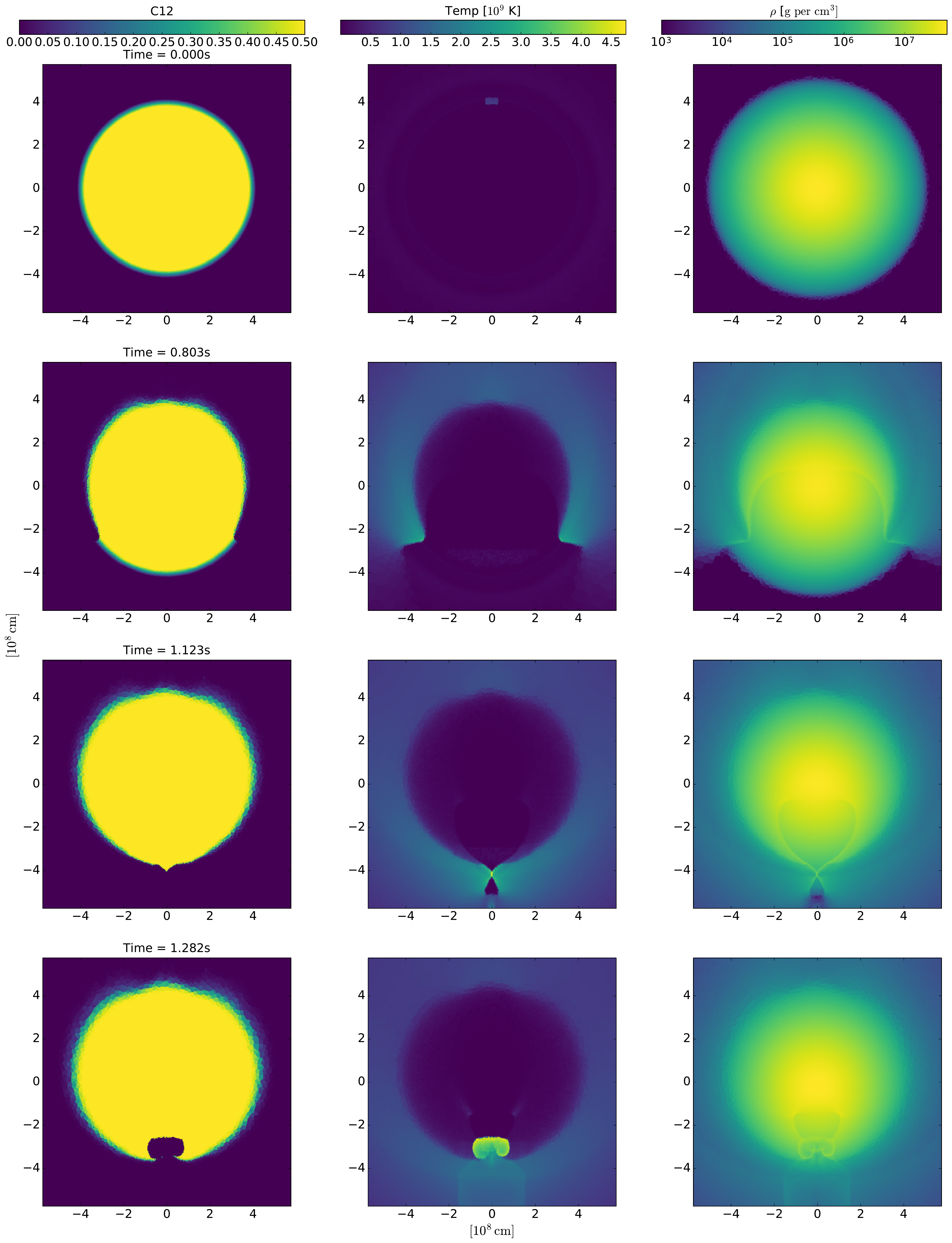}
  \caption{Time evolution of Model M2a; shown from left to right: mass
  fraction of carbon, temperature in K, and density in g cm$^{-3}$; shown
  from top to bottom: at time $t=0$\,s, $t=0.803$\,s, $t=1.123$\,s, and
  $t=1.282$\,s plotted as slices through the center of the white dwarf in the $x\,-\,z$
  plane.}
  \label{fig:mechanism}
\end{figure*}

The propagation of the detonation wave is visible in Fig.~\ref{fig:mechanism}.
It shows the evolution of the carbon mass fraction, temperature, and density
(from left to right) at four different times.

The top row illustrates the ignition of the helium detonation in the high
temperature spot. The location of the core-shell transition can be inferred
from the carbon abundance. The second row shows the propagation of the
detonation wave in the helium shell and the shock in the CO core. The burning
in the helium shell is visible as some of the carbon in the core-shell
transition region is burned together with the helium. At the same time the
energy release leads to a temperature increase. The shock wave driven into the
core by the helium detonation is visible in the density slice. The third row of
Fig.~\ref{fig:mechanism} shows the instant when the detonation wave that
propagated through the He shell converges into a spot on the far side of the
ignition. This initiates a second detonation in the carbon-oxygen core
$1.123$\,s after helium ignition. Details of this detonation formation are
described in Section~\ref{sec:mechanism}. After the convergence of the He
detonation wave and ignition of a carbon detonation, the detonation wave moves
inwards with a velocity of about $13.1\times10^8\text{cm s}^{-1}$ and a
detonation wave develops in the core. The carbon detonation wave incinerates
the core, thereby running over the shock wave still propagating through the
core material. The bottom row in Fig.~\ref{fig:mechanism} illustrates this
situation: The shock wave sent into the core from the shell detonation is
about to converge while the detonation overruns it.

Small asymmetries are visible in the bottom row of Fig.~\ref{fig:mechanism}
which persist for the remaining simulation time until homologous expansion sets
in. These asymmetries might be caused by the irregularity of the grid or the
initial asymmetric initiation of the detonation in the helium shell. The
Voronoi mesh consists of cells that have different sizes and shapes. Only the
cell mass is a constant parameter. This grid does not allow a perfect
representation of all shapes. The characteristic of the mesh also results in
small asymmetries in the initial artificial detonation spot in the helium
shell as discussed in Section~\ref{sec:detonation}. Due to this, the propagation
of the detonation wave in the helium shell might differ slightly between
different sides of the white dwarf.

\subsection{CO detonation ignition mechanism}
\label{sec:mechanism}
The detonation ignition mechanism shown in Fig.~\ref{fig:mechanism} has not been
investigated in detail in previous work, which mostly focused on the converging
shock mechanism as the possible trigger for the second detonation. It has, however,
been briefly described by \citet{livne1995a}, \citet{garcia1999a},
\citet{forcada2007a}, and \citet{garcia2018b}. We call this mechanism the
scissors mechanism because the detonation forms as a shock wave propagates into
dense core material while sliding over each other, much like closing scissors.

\begin{figure*}
  \centering
  \includegraphics[width=1.0\textwidth]{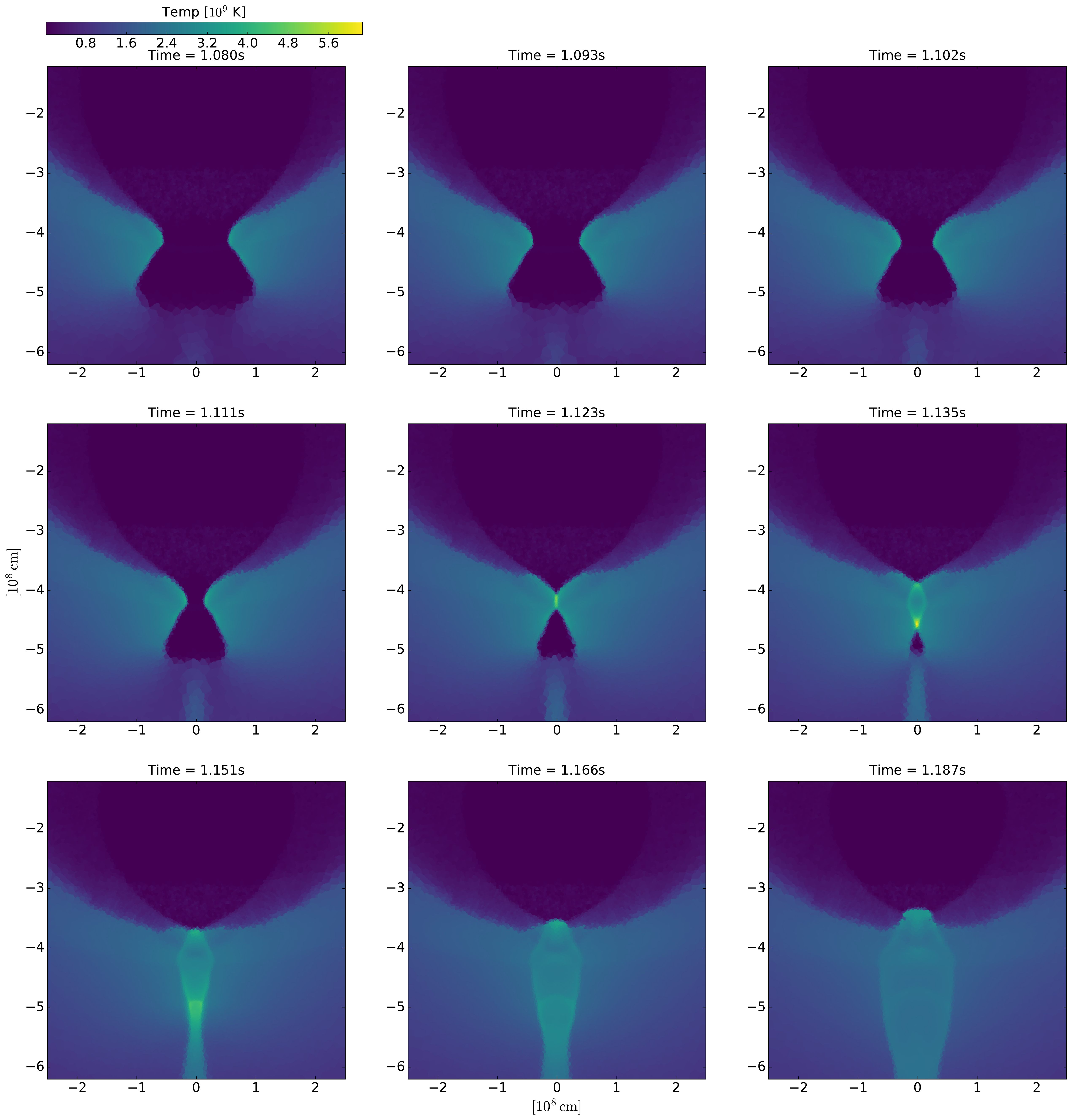}
  \caption{Zoom-in of Fig.~\ref{fig:mechanism} on the time evolution of
  Model M2a; temperature profile with increasing time from left to right and top
  to bottom: the times are indicated above each plot going from $t=1.080$\,s in
  the top left to $t=1.187$\,s in the bottom right; plotted as slices through
  the center of the white dwarf in the $x\,-\,z$ plane.}
  \label{fig:mechanismdet}
\end{figure*}

Fig.~\ref{fig:mechanismdet} shows the temperature in a zoom-in (in time and
space) into the region of $-2.5\times10^8\,\text{cm} < \text{x} <
2.5\times10^8\,\text{cm} $ and $ -6.5\times10^8\,\text{cm} < \text{z} <
-2.5\times10^8\,\text{cm} $ of Fig.~\ref{fig:mechanism} at 1.080\,s to 1.187\,s
after the first detonation ignition. At the helium detonation front the
temperature is high with values of about $3.6\times10^9$\,K. It closes up
1.123\,s after its ignition on the far side of the WD star (see the center
panel of Fig.~\ref{fig:mechanismdet}). A comparison of the profile of the
carbon abundance and temperature in the third row from the top in
Fig.~\ref{fig:mechanism} shows that the point of convergence is close to the
base of the helium shell. This region contains carbon which was mixed into the
shell during the relaxation and acts as fuel that can be burned. Peak
temperatures of about $2.7\times10^9$\,K are reached. Temperature spikes of at
least $2.4\times10^9$\,K lead to explosive burning in a few cells.  These cells
each have a volume of about $3.22\times10^{19}$\,cm$^3$ on average
(corresponding to a radius of about 20\,km assuming a spherical structure of
the cell) with a density higher than $3.0\times10^6\text{g cm}^{-3}$ after
1.123\,s. In all detonating cells, the temperature increases further to above
$2.8\times10^9$\,K 1.126\,s after the first ignition in the helium shell
increasing the robustness of the detonation.  The detonating cells have
abundances of at least $0.20$ in carbon and $0.42$ in oxygen, confirming the
carbon detonation.

Even in our highest-resolution simulation, the ignition of the carbon
detonation is far from being resolved. The detonating cells in the simulation
with the highest resolution have a radial extent of about 20\,km.
\citet{katz2019a} in comparison show that a resolution lower than 1\,km is
necessary which can not be reached easily in full 3D simulations. The
detonation in our simulations is at least in parts due to numerical effects. We
therefore compare with off-line detonation ignition studies to determine
whether a detonation can form physically under the conditions we observe in our
simulations. The values of temperature and density are high enough to trigger
a detonation according to both \citet{roepke2007a} and \citet{seitenzahl2009b}
who determined critical values and sizes for detonation ignition.
\citet{roepke2007a} list a temperature of $2.3\times10^9$\,K to be sufficient
on a 100\,km scale with a density of at least $1.41\times10^6\text{g cm}^{-3}$.
\citet{seitenzahl2009b} give values of at least $5.0\times10^6\text{g cm}^{-3}$
and $2.0\times10^9$\,K for the density and temperature, respectively. We
therefore conclude that a physical detonation ignition at the points singled
out in our simulation is likely to occur.

\subsection{Final abundances}
\label{sec:abundances}
\begin{table*}[htbp]
    \caption{Final abundances in the explosion ejecta for Models M1a,
    FM3$^{\text{(1),(2)}}$, M2a, and M2a\_i55.}
  \label{tab:abund}
  \centering
  \begin{tabular}{@{}lrrrr|rrrr}
  \hline
  & \multicolumn{4}{c}{He detonation} & \multicolumn{4}{c}{core detonation} \\
  & M1a & FM3 & M2a & M2a\_i55 & M1a & FM3 & M2a & M2a\_i55 \\
  & [M$_\odot$] & [M$_\odot$] & [M$_\odot$] & [M$_\odot$] & [M$_\odot$] & [M$_\odot$] & [M$_\odot$] & [M$_\odot$] \\ \hline
  $^4$He & $2.5\times10^{-2}$ & $3.3\times10^{-2}$ & $2.3\times10^{-2}$ &$2.3\times10^{-2}$ & $4.2\times10^{-3}$ & & $5.0\times10^{-3}$ & $5.4\times10^{-3}$\\
  $^{12}$C & $3.6\times10^{-4}$ & $2.2\times10^{-4}$ & $1.0\times10^{-4}$ & $6.8\times10^{-5}$ & $1.2\times10^{-3}$ & $2.7\times10^{-3}$ & $8.9\times10^{-4}$ & $8.2\times10^{-4}$\\
  $^{16}$O & $5.0\times10^{-3}$ & $1.9\times10^{-6}$ & $7.4\times10^{-3}$ & $7.6\times10^{-3}$ & $5.5\times10^{-2}$ & $8.0\times10^{-2}$ & $5.2\times10^{-2}$ & $5.2\times10^{-2}$\\
  $^{28}$Si & $4.6\times10^{-3}$ & $1.4\times10^{-4}$ & $8.9\times10^{-3}$ & $9.1\times10^{-3}$ & $1.7\times10^{-1}$ & $2.1\times10^{-1}$ & $1.6\times10^{-1}$ & $1.5\times10^{-1}$\\
  $^{32}$S & $1.8\times10^{-3}$ & $7.8\times10^{-4}$ & $3.2\times10^{-3}$ & $3.3\times10^{-3}$ & $1.1\times10^{-1}$ & $1.0\times10^{-1}$ & $1.1\times10^{-1}$ & $1.0\times10^{-1}$\\
  $^{40}$Ca & $2.7\times10^{-3}$ & $2.2\times10^{-3}$ & $3.6\times10^{-3}$ & $3.5\times10^{-3}$ & $2.4\times10^{-2}$ & $1.8\times10^{-2}$ & $2.3\times10^{-2}$ & $2.2\times10^{-2}$\\
  $^{44}$Ti & $7.2\times10^{-4}$ & $3.4\times10^{-3}$ & $7.0\times10^{-4}$ & $6.9\times10^{-4}$ & $2.8\times10^{-5}$ & $1.1\times10^{-5}$ & $2.8\times10^{-5}$ & $2.9\times10^{-5}$\\
  $^{48}$Cr & $1.5\times10^{-3}$ & $4.4\times10^{-3}$ & $1.6\times10^{-3}$ & $1.6\times10^{-3}$ & $4.9\times10^{-4}$ & $4.5\times10^{-4}$ & $4.8\times10^{-4}$ & $4.7\times10^{-4}$ \\
  $^{56}$Ni & $1.5\times10^{-2}$ & $1.7\times10^{-3}$ & $1.2\times10^{-2}$ & $1.2\times10^{-2}$ & $5.6\times10^{-1}$ & $5.5\times10^{-1}$ & $5.7\times10^{-1}$ & $5.9\times10^{-1}$ \\ \hline
\end{tabular}
  \tablebib{
  (1)~\citet{fink2010a}, (2)~\citet{kromer2010a}}
\end{table*}

\noindent
The final abundances of $^4$He, $^{12}$C, $^{16}$O, $^{28}$Si, $^{32}$S,
$^{40}$Ca, $^{44}$Ti, and $^{56}$Ni in the ejected material as determined from
nucleosynthetic postprocessing of the tracer particles (see
Section~\ref{sec:theory_pp}) are given in Table~\ref{tab:abund} for Models
M1a, FM3, M2a, and M2a\_i55. The four left columns list the abundances
from the helium detonation, while the four right columns give the final
abundances of the core detonation when homologous expansion occurs. The tracer
particle approach allows us to separate the yields of the helium detonation
from the yields of the core detonation: we identify tracers that are
affected by the helium shell detonation based on whether it had a helium
mass fraction of at least 0.01 at the beginning of the hydrodynamic detonation
simulation. \citet{fink2010a} do not consider the possibility of the ignition
of a detonation at the point where the detonation wave converges at the
shell-core interface for Model FM3, but they focus on a second detonation that
might be ignited in the CO core due to spherical shock collimation near the
center. Their abundances are included in Table~\ref{tab:abund} for comparison
and taken from \citet{fink2010a} and \citet{kromer2010a} for Model FM3.

In Model M2a about equal parts of $^{32}$S and $^{40}$Ca are produced
in the He shell detonation ($\sim$\,$10^{-3}$\,M$_\odot$) while the amounts of
produced $^{16}$O and $^{28}$Si are slightly higher. The $^{44}$Ti abundance is
comparatively low with $7.0\times10^{-4}$\,M$_\odot$. The $^{12}$C is
also low -- in this case because high enough temperatures are reached to burn
to heavier elements. The relatively high amount of helium in the shell
detonation is a result of the expansion of the matter during the burning. The
density decreases, it cools down and burning stops.

The high temperatures and densities together with carbon in the helium shell
allow the production of more heavy elements such as $^{56}$Ni compared to Model
FM3. The $\alpha$-captures are accelerated due to the presence of $^{12}$C (see
\citet{fink2010a} for a detailed discussion). There is a difference of about
one order of magnitude in most species produced in the helium detonation
between our reference model M2a and FM3. Only the $^{12}$C and
$^{40}$Ca abundances are about the same with
$1.0\times10^{-4}$\,M$_\odot$ and $2.2\times10^{-4}$\,M$_\odot$, and
$3.6\times10^{-3}$\,M$_\odot$ and $2.2\times10^{-3}$\,M$_\odot$, respectively.
The $^{44}$Ti abundance shows a reduction by about a factor of five
compared to the value found by \citet{fink2010a}. In Model M2a we burn
about 50\,\% of the initial helium. This is higher than in \citet{fink2010a}
and can in part explain our higher $^{56}$Ni production.  Another reason for
the discrepancy is the different modeling approach of \citet{fink2010a}. Their
level set method does not allow a self-consistent calculation of the energy
release and nuclear burning.\\

The abundances of the core detonation are in agreement with the qualitatively
expected values. $^{56}$Ni is produced most abundantly followed by $^{28}$Si
and $^{32}$S. The mixing of carbon into the shell causes less carbon to be
present in the core so that less is present after the detonation. The
abundances of the core detonation are of the same order in Models M2a and FM3,
except for the amount of $^{12}$C which is about one order of magnitude lower
in M2a than in FM3 after 100\,s in our simulation. Small variations in the
abundances of the core detonation between M2a and FM3 can be explained by the
different codes. The agreement of both models is better for the core detonation
than the helium detonation as the core mass and composition are very similar.
Moreover, the level set method to model detonations is more precise at the high
densities in the core. The total final mass (shell and core) of $^{56}$Ni
($0.59M_\odot$) is in the expected range for a normal Type Ia supernova
\citep{stritzinger2006b,scalzo2014a}. A discussion on how the change in the
abundances influences the spectra follows in Section~\ref{sec:RT}.

\subsection{Influence of core and shell mixing}
\label{sec:mixing}

\begin{figure}[h]
    \includegraphics[width=0.42\textwidth]{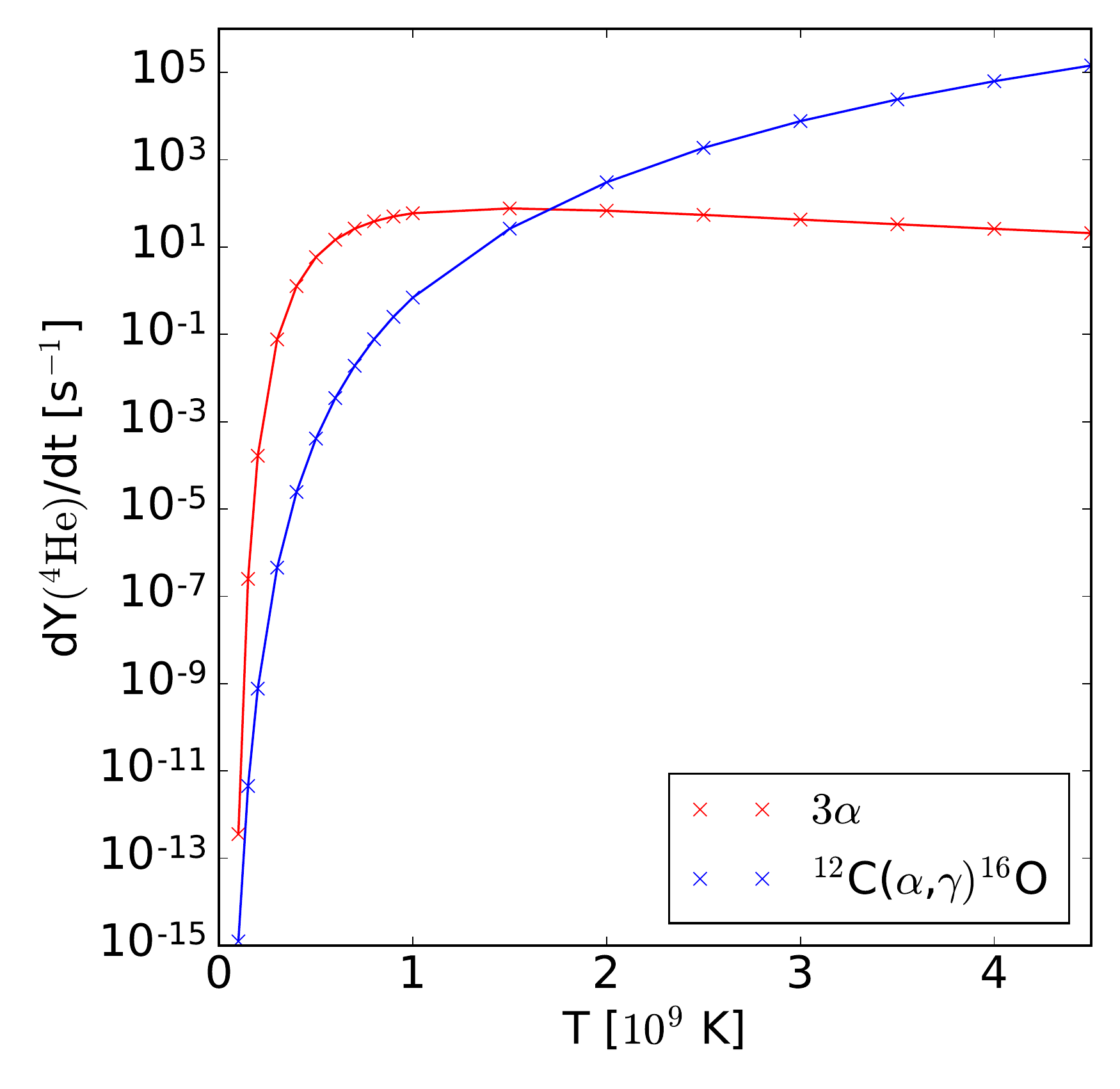}
    \includegraphics[width=0.44\textwidth]{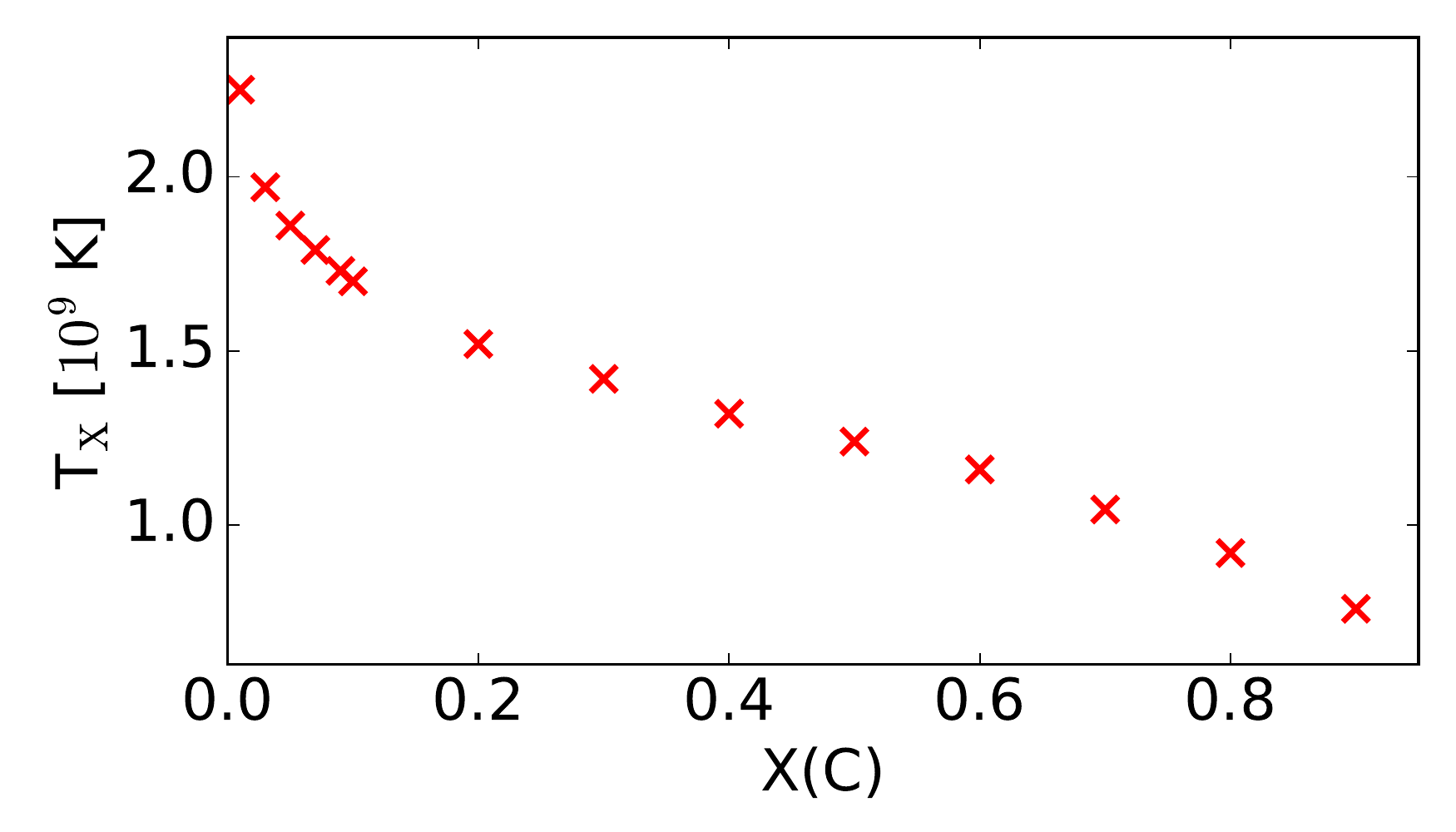}
    \caption{Top: Rate of change in $^4$He abundance due to the
        triple-$\alpha$ (red) and $^{12}$C$(\alpha, \gamma)$$^{16}$O (blue)
        reactions at a density of $1.2~\times~10^{6}~\,~\mathrm{g}\,
        \mathrm{cm}^{-3}$ and $X(^{12}\mathrm{C}) = 0.1$ dependent on
        temperature; bottom: crossing temperature $T_{\text{X}}$
            above which the depletion due to $\alpha$-captures exceeds that
            due to triple-$\alpha$ dependent on initial carbon abundance at a
            density of $1.2 \times 10^{6} \, \mathrm{g}\, \mathrm{cm}^{-3}$.
    The reaction rates are taken from the JINA Reaclib Database
    \citep{cyburt2010a} based on \citet{xu2013a} for the $\alpha$-capture and
    \citet{fynbo2005a} for the triple-$\alpha$ reaction.}
    \label{fig:Cenrichment}
\end{figure}

Until now it is unclear whether the transition between core and shell is sharp
or smeared out over a certain volume. \mbox{\citet{neunteufel2017a}} expect that some
-- but not very much -- mixing takes place during the accretion of helium onto
the white dwarf. The ignition of He shell burning may dredge up some core
material. Even without mixing, the material forming the He shell has some
metallicity. Accretion from a hybrid HeCO WD is another possibility. An
enrichment with carbon (and possibly other metals) influences the He burning.
As a simple illustration, we consider the case of material consisting of $^4$He
and $^{12}$C only. Similar arguments can be made for the admixture of other
metals \citep[see][]{shen2014b}. Carbon admixture in the He shell has two
effects: it enhances the burning rate and can limit the mass number of the ash
material. \citet{kromer2010a} show how this impacts synthetic spectra and light
curves and conclude that it leads to a better fit with the observations for
Model FM3.

The triple-$\alpha$ reaction is the bottle-neck in the pure He
composition case. In a pure He shell the triple-$\alpha$ reaction
first needs to produce $^{12}$C before the $\alpha$-process starting
out with $^{12}$C$(\alpha, \gamma)$$^{16}$O forms heavier elements up
to $^{56}$Ni. A seed abundance of carbon circumvents the
triple-$\alpha$ bottleneck. This is because for the temperatures
reached after ignition of explosive helium burning the
$\alpha$-capture is faster than the triple-$\alpha$ reaction (see top
panel of Fig.~\ref{fig:Cenrichment}). Such an effect occurs for
any carbon mass fraction above some cross-over temperature
$T_\mathrm{X}$ (Fig.~\ref{fig:Cenrichment}, bottom panel). Since
$T_\mathrm{X}$ is typically smaller than the temperatures reached
in explosive He burning, this enhances the burning rate in the He
shell and leads to stronger shocks by increasing the energy release in
the detonation.

The second effect in addition to this carbon-enhanced He detonation
results from an over-pollution of the He shell with carbon: In the
$\alpha$-process, it takes eleven $\alpha$ particles to reach $^{56}$Ni from
$^{12}$C. Therefore, for a number ratio of helium to carbon smaller than
$11$:$1$ we enter the $\alpha$-limited regime, where the $\alpha$-chain
stalls around a nucleus with nucleon number $A < 56$.  The stagnation point of
the $\alpha$-chain for a given carbon mass fraction is determined by the
relations
\begin{equation}
  \label{eq:1}
12+4n = A,
\end{equation}
where $n$ is the number of $\alpha$ particles needed to reach the stagnation
nucleus from $^{12}$C, and
\begin{equation}
  \label{eq:2}
  \frac{Y(^{4}\mathrm{He})}{Y(^{12}\mathrm{C})} = n = 3\frac{X(^{4}\mathrm{He})}{X(^{12}\mathrm{C})}.
\end{equation}
Since, by mass conservation of a two species mixture,
\begin{equation*}
X(^{4}\mathrm{He}) = 1 - X(^{12}\mathrm{C}),
\end{equation*}
Eq.~(\ref{eq:2}) gives $n = 3/X(^{12}\mathrm{C}) - 3$. Substituting n into
Eq.~(\ref{eq:1}) we finally get the nucleon number of the stagnation nucleus
\begin{equation*}
A = \frac{12}{X(^{12}\mathrm{C})}.
\end{equation*}
The $\alpha$-limited regime thus begins at a mass fraction X(C) > 0.21. In
their Model 3m, \citet{kromer2010a} add 34\% by mass of $^{12}$C to the He
shell and therefore reach a stagnation of the $\alpha$~-~process around argon
which avoids strong imprints on the predicted optical spectra of the simulated
supernova explosion. While \cite{kromer2010a} study a homogeneous admixture
of carbon to the shell, our models are characterized by an abundance profile
with a higher amount of carbon at the base of the shell than at its outer edge
that results from the relaxation process (see Section \ref{sec:relax}). This
gradient represents the expected abundance distribution more realistically
\citep{neunteufel2017a}.

To bracket the effect and to investigate the influence of carbon in the shell
on the detonation ignition mechanism, a different structure in the helium,
carbon, and oxygen abundances is modeled. For this, no additional mixing of
carbon and oxygen into the He shell is added after the relaxation in Model M1a:
In this model the core consists of pure carbon and oxygen in equal mass parts
and the helium shell is only slightly enriched with carbon. As a result of this
only about $0.007\,M_\odot$ of each carbon and oxygen is present in the shell
and the total shell mass decreases. The change in the composition of the
transition region has an influence on the carbon detonation ignition mechanism.
Contrary to Model M2a, the convergence of the helium detonation wave
at the antipode is not strong enough to ignite a carbon detonation at the
shell-core interface in Model M1a. However, a detonation according to
the converging shock mechanism is found.  Moreover, the omission of additional
mixing in Model M1a results in small differences of the final abundances
compared to those of Model M2a. Because of the similar shell mass and
detonation ignition mechanism the abundances of Model M1a can best be compared
to Model FM3 (see Table~\ref{tab:abund}). The small addition of carbon to the
shell in the transition region leads to an increased production of IMEs
compared to FM3 while one order of magnitude less $^{44}$Ti and more $^{56}$Ni
is produced in the shell detonation. These differences can be explained by the
different treatment of the helium shell detonation in both models.\\ Models M1a
and 3m by \cite{kromer2010a} lead us to conclude that the mixing of carbon to
the helium shell has an impact on the yields as it results in the production of
heavier elements. However, our models show that it does not solve the problem
of a significant amount of $^{4}$He being unburned and the redness of the
synthetic spectra \citep{boyle2017a,botyanszki2018a}. On the other hand the
mixing is critical for the details of the detonation ignition mechanism as
stated above.

\subsection{Resolution study}
\label{sec:resolution}

\begin{table*}[h]
    \caption{Reference mass $M_\text{R}$ of the helium shell and at the
    carbon ignition point as well as energy release of the He shell detonation
    of Models M2a, M2a\_13, M2a\_21, M2a\_36, and M2a\_79. The last column lists
    whether a carbon detonation ignition following the scissors mechanism is
    observed.}
  \label{tab:resolution}
  \centering
  \begin{tabular}{@{}lrrrr}
    \hline
    & $M_\text{R}$ in He shell & E$_{\text{He shell}}$ & $M_\text{R}$ at C ign. point & scissors mechanism \\
    & [$10^{26}$\,g]  & [erg] & [$10^{26}$\,g] &  \\ \hline
    M2a\_79 & 20.0 & $9.78\times10^{49}$ & &  \\
    M2a\_36 & 4.0 & $9.93\times10^{49}$ & & \\
    M2a\_21 & 2.0 & $9.97\times10^{49}$ & 2.0 & no \\
    M2a\_13 & 2.0 & & 1.2 & yes \\
    M2a     & 2.0 & & 0.2 & yes \\ \hline
  \end{tabular}
\end{table*}

In this study, we are interested in the results of the helium shell detonation
and the mechanism of the core detonation ignition. We therefore perform
numerical convergence studies for the two effects in separate steps. Models
M2a\_79, M2a\_36, and M2a\_21 have different refinements in the helium shell,
characterized by the reference mass $M_\text{R}$ as given in the first column
of Tab.~\ref{tab:resolution}. The total number of cells 1.123\,s after helium
ignition is 2.0\,million, 1.6\,million, and 1.3\,million for Models M2a\_21,
M2a\_36, and M2a\_79. The energy release of the He shell detonation (see second
column of Tab.~\ref{tab:resolution}) shows convergence: the difference of the
energy release between the simulation with the high (M2a\_21) and modest
(M2a\_36) resolution is smaller than the difference between M2a\_36 and
M2a\_79.

Model M2a\_21, which has the highest He shell refinement, is taken as the base
for testing the convergence of the carbon detonation ignition mechanism.  Two
further simulations have an additional refinement around the carbon ignition
point (see third column in Tab.~\ref{tab:resolution}). A detonation ignition by
the scissors mechanism is not observed in Model M2a\_21, but for both models
with higher resolution in the corresponding material. This indicates that core
detonation ignition by the scissors mechanism is the converged numerical
solution.

\subsection{Sensitivity to the ignition spot}
As described in Section \ref{sec:detonation} the ignition spot for the first
detonation was changed to consider two different locations in Model M2. The
simulations M2a and M2b differ only slightly (see Table~\ref{tab:models}). The same
propagation behavior is observed when the ignition spot is set to be located at the
base of the He shell (Model M2b) and a carbon detonation is ignited at the same
location as in Model M2a. The detonation ignition mechanism is robust against small
changes in the location of the first ignition spot of the helium detonation.

\subsection{Influence of a different white dwarf mass}
\label{sec:comparison}
\begin{table}[htbp]
    \caption{Final abundances for Model M3a and FM1$^{\text{(1),(2)}}$.}
  \label{tab:M2babund}
  \centering
  \begin{tabular}{@{}lrr|rr}
  \hline
  & \multicolumn{2}{c}{He detonation} & \multicolumn{2}{c}{core detonation} \\
  & M3a & FM1 & M3a & FM1 \\
  & [M$_\odot$] & [M$_\odot$] & [M$_\odot$] & [M$_\odot$] \\ \hline
  $^4$He & $4.2\times10^{-2}$ & $8.3\times10^{-2}$ & $1.4\times10^{-3}$ & \\
  $^{12}$C & $7.6\times10^{-5}$ & $1.2\times10^{-3}$ & $4.0\times10^{-4}$ & $6.6\times10^{-3}$\\
  $^{16}$O & $1.7\times10^{-2}$ & $3.2\times10^{-6}$ & $6.8\times10^{-2}$ & $1.4\times10^{-1}$\\
  $^{28}$Si & $2.7\times10^{-2}$ & $4.8\times10^{-4}$ & $1.8\times10^{-1}$ & $2.7\times10^{-1}$\\
  $^{32}$S & $5.0\times10^{-3}$ & $2.2\times10^{-3}$ & $1.2\times10^{-1}$ & $1.3\times10^{-1}$\\
  $^{40}$Ca & $4.2\times10^{-3}$ & $4.7\times10^{-3}$ & $2.3\times10^{-2}$ & $2.0\times10^{-2}$\\
  $^{44}$Ti & $1.3\times10^{-3}$ & $7.9\times10^{-3}$ & $1.9\times10^{-5}$ & $7.2\times10^{-6}$\\
  $^{48}$Cr & $2.5\times10^{-3}$ & $1.1\times10^{-2}$ & $4.4\times10^{-4}$ & $3.9\times10^{-4}$\\
  $^{56}$Ni & $3.1\times10^{-2}$ & $8.4\times10^{-4}$ & $3.1\times10^{-1}$ & $1.7\times10^{-1}$\\ \hline
  \end{tabular}
\tablebib{
    (1)~\citet{fink2010a}, (2)~\citet{kromer2010a}}
\end{table}

A comparison of Model M2a to Model M3a is carried out to consider the effects
different white dwarf masses might have on the mechanism.  Details of Model M3a
are listed in Table~\ref{tab:models}. The size of the helium detonation is
slightly smaller than in Model M2a.  2447~cells are ignited with $\Delta R$ set
to be $0.02$ times the distance between central detonating cell and the center
of the white dwarf as described in Sec.~\ref{sec:detonation}. The total mass of
the white dwarf here is about 0.9\,M$_\odot$ with a helium shell about twice as
massive as in Model M2a. The simulation is run at a high resolution and shows a
convergence of the detonation wave 1.251\,s after the first ignition.  At this
point, opposite to the helium detonation spot, a density of at least
$5.8\times10^6$\,g cm$^{-3}$ and temperature higher than $3.2\times10^9$\,K is
reached in cells with a mass fraction of $0.32$ in $^{12}$C leading to the same
detonation ignition mechanism as in Model M2a.

The final abundances gained from Model M3a are listed in
Table~\ref{tab:M2babund}. A total of 0.34\,M$_\odot$ of $^{56}$Ni is produced
in the shell and core detonations combined. The total yields for $^{28}$Si and
$^{32}$S are produced most abundantly following $^{56}$Ni in this simulation
with 0.21\,M$_\odot$ and 0.12\,M$_\odot$, respectively. The final abundances of
Model FM1 are included in Table~\ref{tab:M2babund} for comparison. Due to the
relaxation process and mixing of carbon and oxygen into the shell in our
simulation the He shell is about 0.03\,M$_\odot$ heavier than in Model FM1.
This leads to a much higher production of $^{56}$Ni and IMEs during the He
detonation, but lower yields of $^{48}$Cr. The abundances from the core
detonation show the same relations as Models M2a and FM3 for most isotopes.
Only the differences in the $^{16}$O and $^{56}$Ni abundances are higher as
Model M3a burns more oxygen and produces twice as much nickel as the model by
\citet{fink2010a}. We remind the reader that \citet{fink2010a} uses a level set
method which is different from our numerical treatment. Their approach is
better suited for a simulation of the WD core at high densities than that of
the helium shell detonation, as discussed in Sec.~\ref{sec:abundances}.

\subsection{Influence of the nuclear network}
We test the sensitivity of the results to the size of the employed nuclear
network. For this we consider a 55-isotope nuclear network during the
hydrodynamics simulation. It is chosen to match the nuclear network in
\citet{townsley2019a} consisting of n, p, $^4$He, $^{11}$B, $^{12-13}$C,
$^{13-15}$N, $^{15-17}$O, $^{18}$F, $^{19-22}$Ne, $^{22-23}$Na, $^{23-26}$Mg,
$^{25-27}$Al, $^{28-30}$Si, $^{29-31}$P, $^{31-33}$S, $^{33-35}$Cl,
$^{36-39}$Ar, $^{39}$K, $^{40}$Ca, $^{43}$Sc, $^{44}$Ti, $^{47}$V, $^{48}$Cr,
$^{51}$Mn, $^{52,56}$Fe, $^{55}$Co, and $^{56,58-59}$Ni. \citet{shen2014b}
point out that a large nuclear network is needed to model the nuclear energy
release accurately. Following this, \citet{townsley2019a} show that a 55
isotope nuclear network is large enough. They argue that their isotope network
gives the same energy release as a large 495-isotope network. The nuclear
network of \citet{townsley2019a} best captures He burning at low
densities while the 33 isotope nuclear network used in M2a is optimized to
follow carbon burning. As our models do not consider a helium shell
enriched with $^{14}$N, there is no need to include a larger nuclear network
similar to the one by \citet{townsley2019a}.

In our Model M2a\_i55, considering the 55 isotope nuclear network, the
second detonation is ignited in the same way as in Model M2a though
the ignition occurs 0.003\,s later. We further find that the total
energy release of the shell and core detonation are within a few per cent of
each other for Models M2a and M2a\_i55 (see Tab.~\ref{tab:energy}).
The final abundances are included in Table~\ref{tab:abund} and it is shown that
the different nuclear network does not change the abundances significantly.
\begin{table}[htbp]
    \caption{Energy release of Models M2a and M2a\_i55.}
  \label{tab:energy}
  \centering
  \begin{tabular}{@{}lrr}
  \hline
  & He detonation & core detonation \\
  & [erg] & [erg] \\ \hline
    M2a & $9.93\times10^{49}$ & $1.35\times10^{51}$ \\
    M2a\_i55 & $1.01\times10^{50}$ & $1.34\times10^{51}$ \\ \hline
  \end{tabular}
\end{table}

\subsection{Comparison to previous work}
\citet{livne1995a}, \citet{garcia1999a}, \citet{forcada2006a}, and
\citet{forcada2007a} found a similar carbon detonation ignition mechanism in their
models. \citet{livne1995a} and \citet{garcia1999a} consider different masses in
their work. \citet{garcia1999a} simulate the explosion of a white dwarf with a
total mass of $1.02$\,M$_\odot$ and \citet{livne1995a} look into different
total masses between $0.70$\,M$_\odot$ and $1.10$\,M$_\odot$. Differences in
the setup, namely the core and shell masses, do not allow us to compare the
final abundances of the models with our work. The model of \citet{forcada2006a}
and \citet{forcada2007a} has a total mass of $0.9$\,M$_\odot$. Their helium
shell is more massive ($0.2$\,M$_\odot$) than our Model M3a which makes a
comparison difficult. Simulations by \citet{garcia2018b} show the same
detonation ignition mechanism. However, they consider different masses for
their models and look into the effect of rotation on the detonation mechanism.

The different models confirm that the detonation ignition mechanism is not
limited to one specific setup as shown in Section \ref{sec:comparison}.
\citet{forcada2007a} finds that the location of the He ignition is important
for the success of the mechanism. Three of their models show a direct carbon
detonation ignition at the surface of the core matching the edge-lit mechanism.
The success of the scissors mechanism may also depend on the thickness of the
transition region between core and shell. Our simulations do not show a prompt
edge-lit detonation \citep[as for example][]{forcada2007a}. \citet{garcia2018b}
find that the convergence of the He detonation waves weakens under the effect
of rotation. Nevertheless, an ignition of the core is still observed at a later
point.

Unlike \citet{livne1995a}, \citet{garcia1999a}, \citet{forcada2007a}, and
\citet{garcia2018b}, work by for instance \citet{fink2007a} does not investigate the
possibility of a mechanism where the convergence of the detonation wave in the
helium shell leads to a second detonation. This is also due to the fact that
they use the level set method to track detonations which prevents an  automatic
ignition of the second detonation, but the core detonation has to be ignited by
hand. They therefore state that their results confirm that a core detonation
would ignite in the converging shock mechanism in case a detonation is not
triggered already at the edge of the core.

\section{Synthetic observables}
\label{sec:RT}

To explore the observable consequences of the scissors mechanism, we have
performed radiative transfer calculations for our reference model M2a.
We selected this model for particular study since it has brightness appropriate
for a typical Type Ia supernova and can be closely compared to results from the
converging shock double-detonation Model FM3.

We have also computed synthetic observables for Models M1a and M2a\_i55.
These allow us to quantify the effect of mixing in Model M2a compared
to Model M1a, and also the effect of using a 33 isotope nuclear
network, compared to Model M2a\_i55 which used a 55 isotope nuclear
network. We reserve for future studies a broader, systematic study of a range
of combined CO and He progenitor masses (models such as M3a). We
recalculated the synthetic observables for Model FM3 in order to ensure that any
differences are due to the explosion models, and not differences in the setup
of our radiative transfer calculations.

In the following we first compare the angle-averaged light curves and spectra
between the models, and comment on their comparison to observations. We then
discuss the viewing angle effects for our synthetic observables.

\subsection{Angle averaged light curves}
\begin{figure*}[h]
\includegraphics[width=0.32\textwidth]{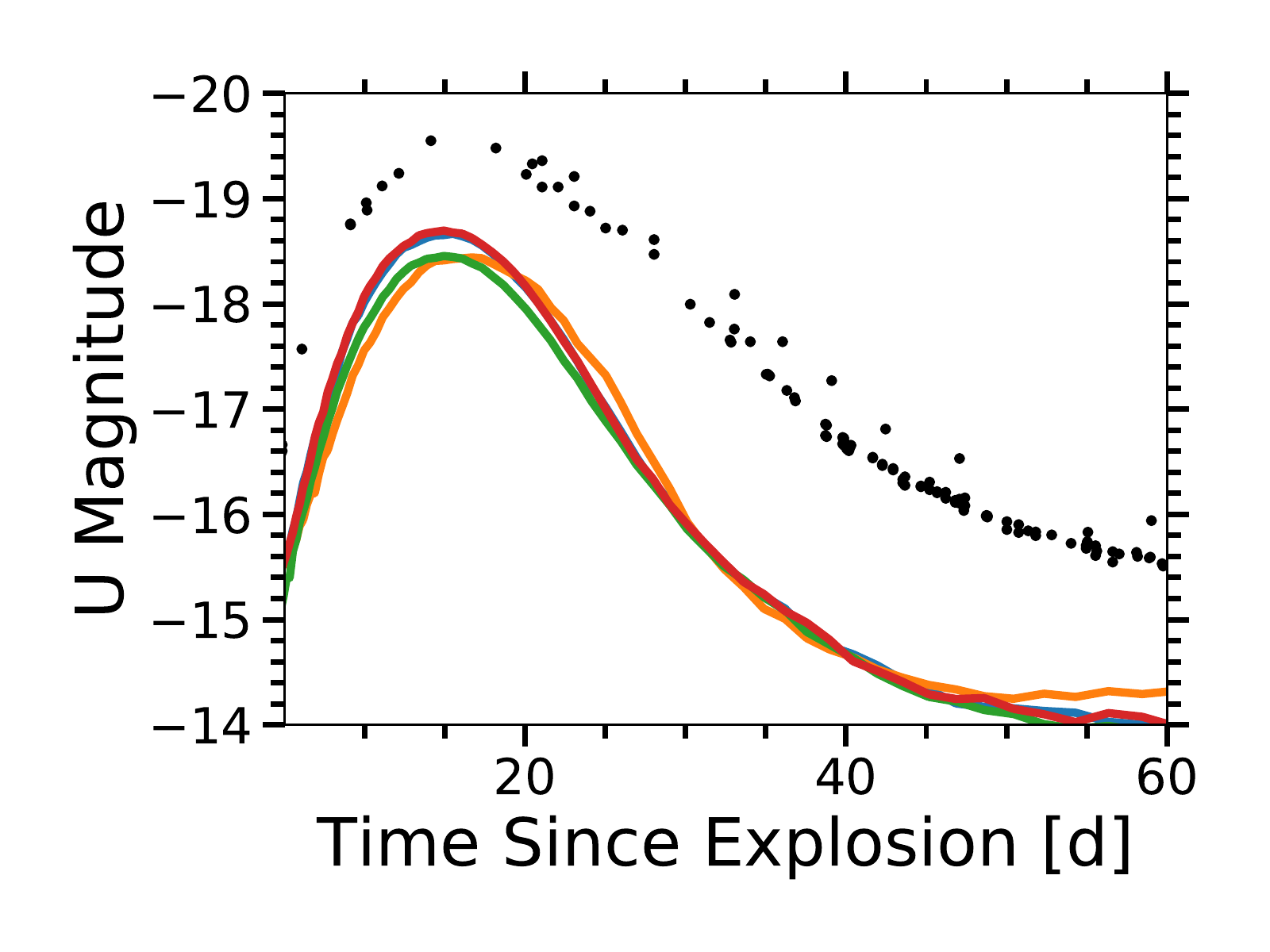}\includegraphics[width=0.32\textwidth]{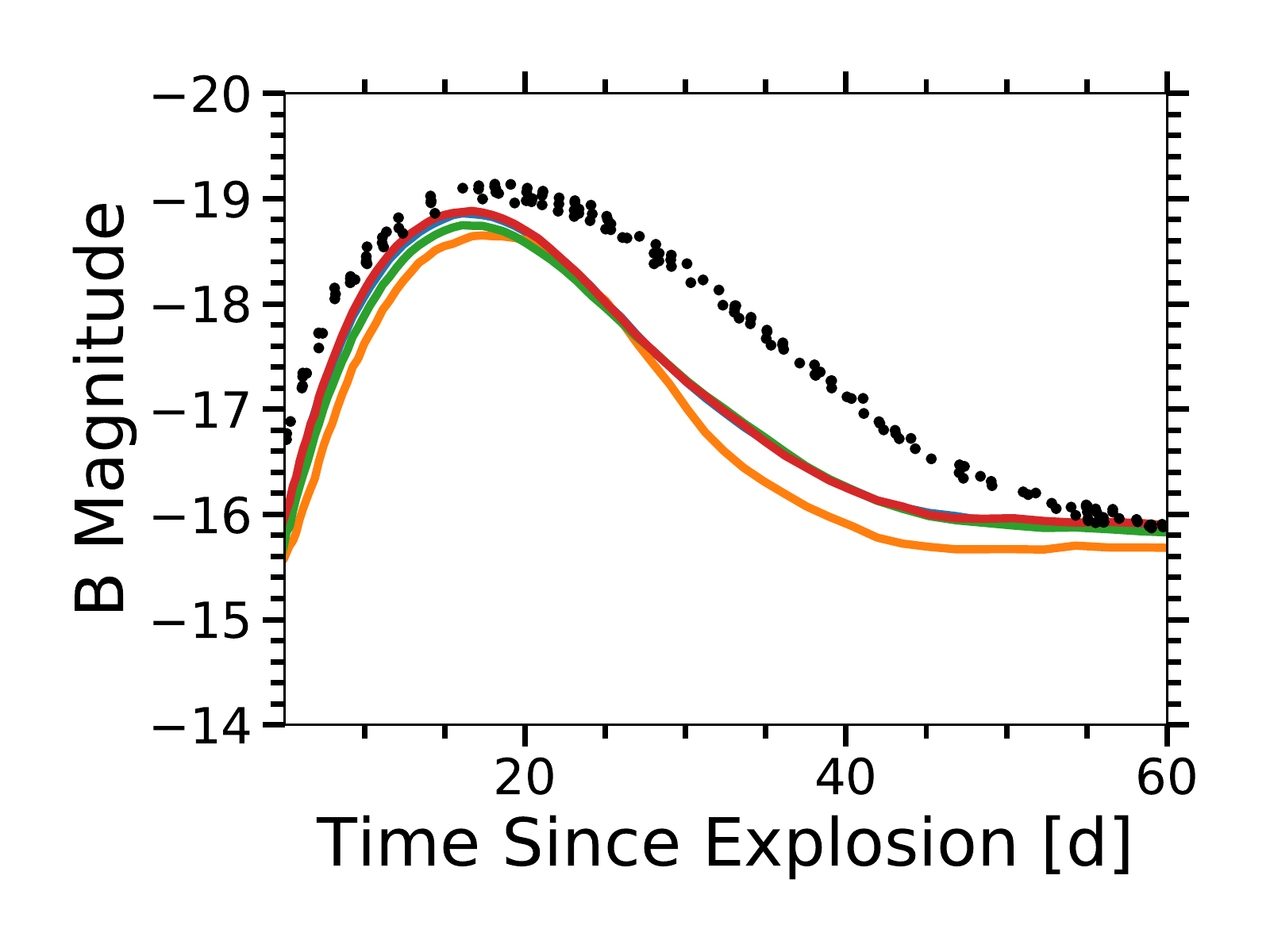}\includegraphics[width=0.32\textwidth]{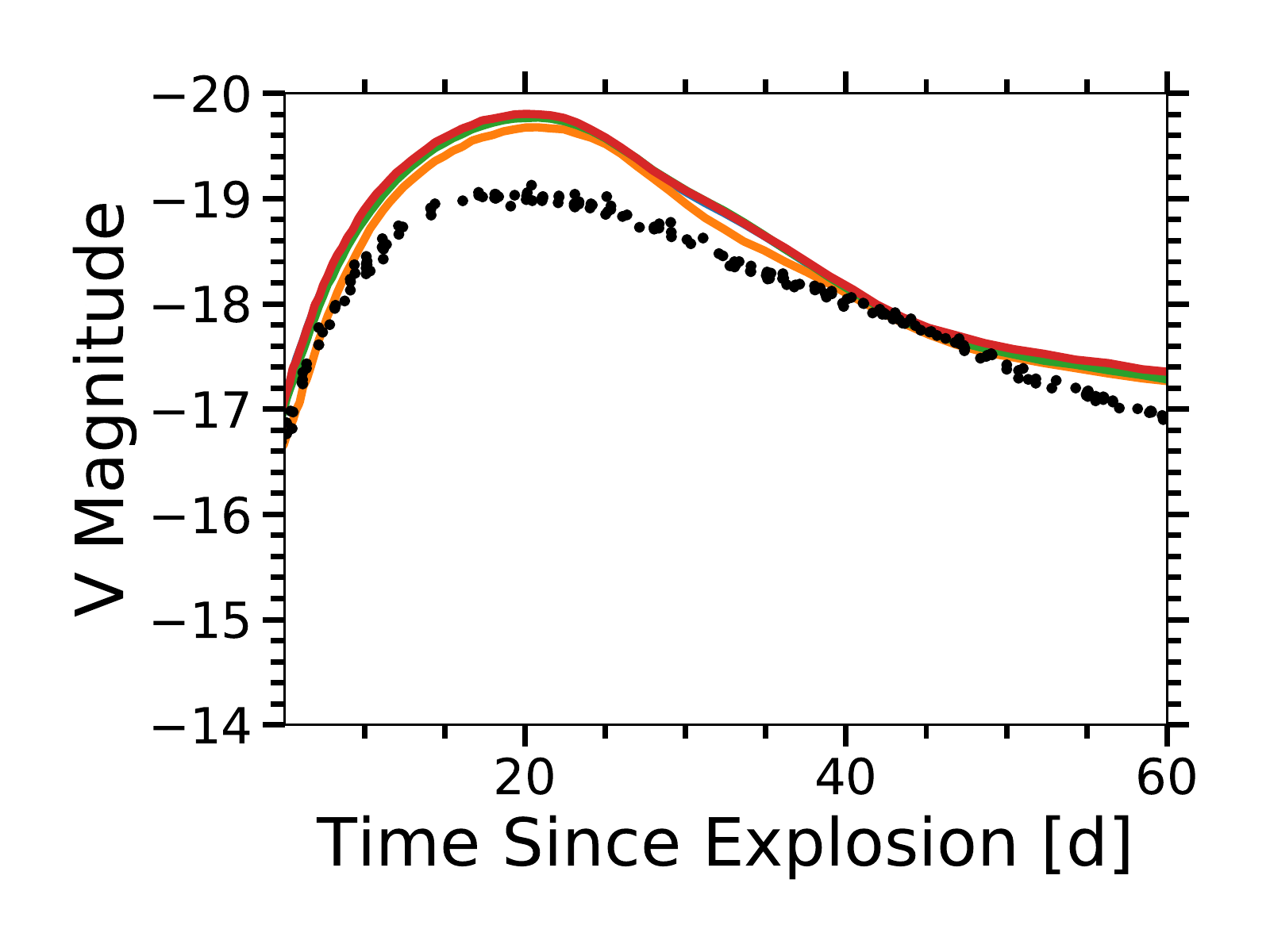}

\includegraphics[width=0.32\textwidth]{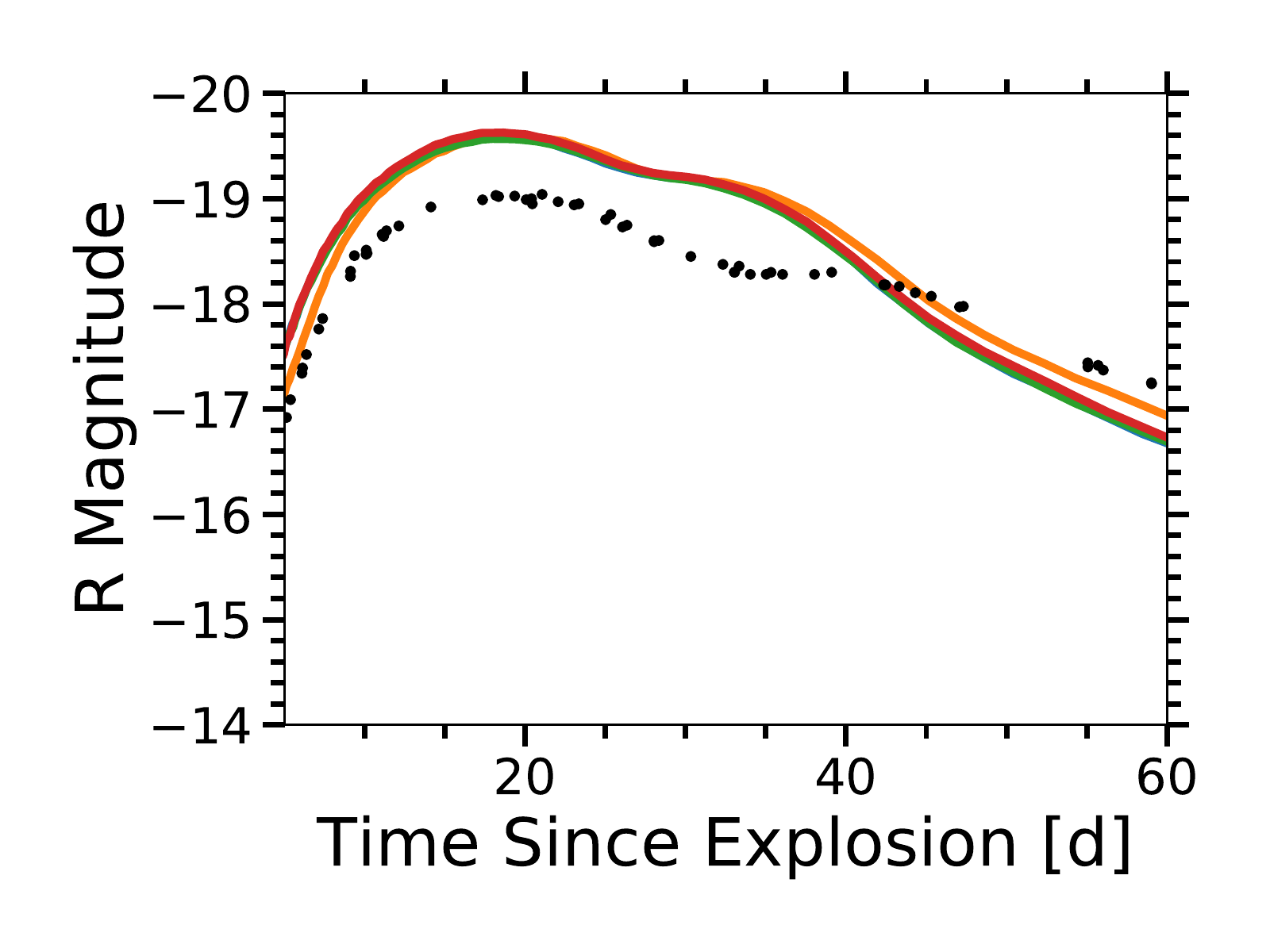}\includegraphics[width=0.32\textwidth]{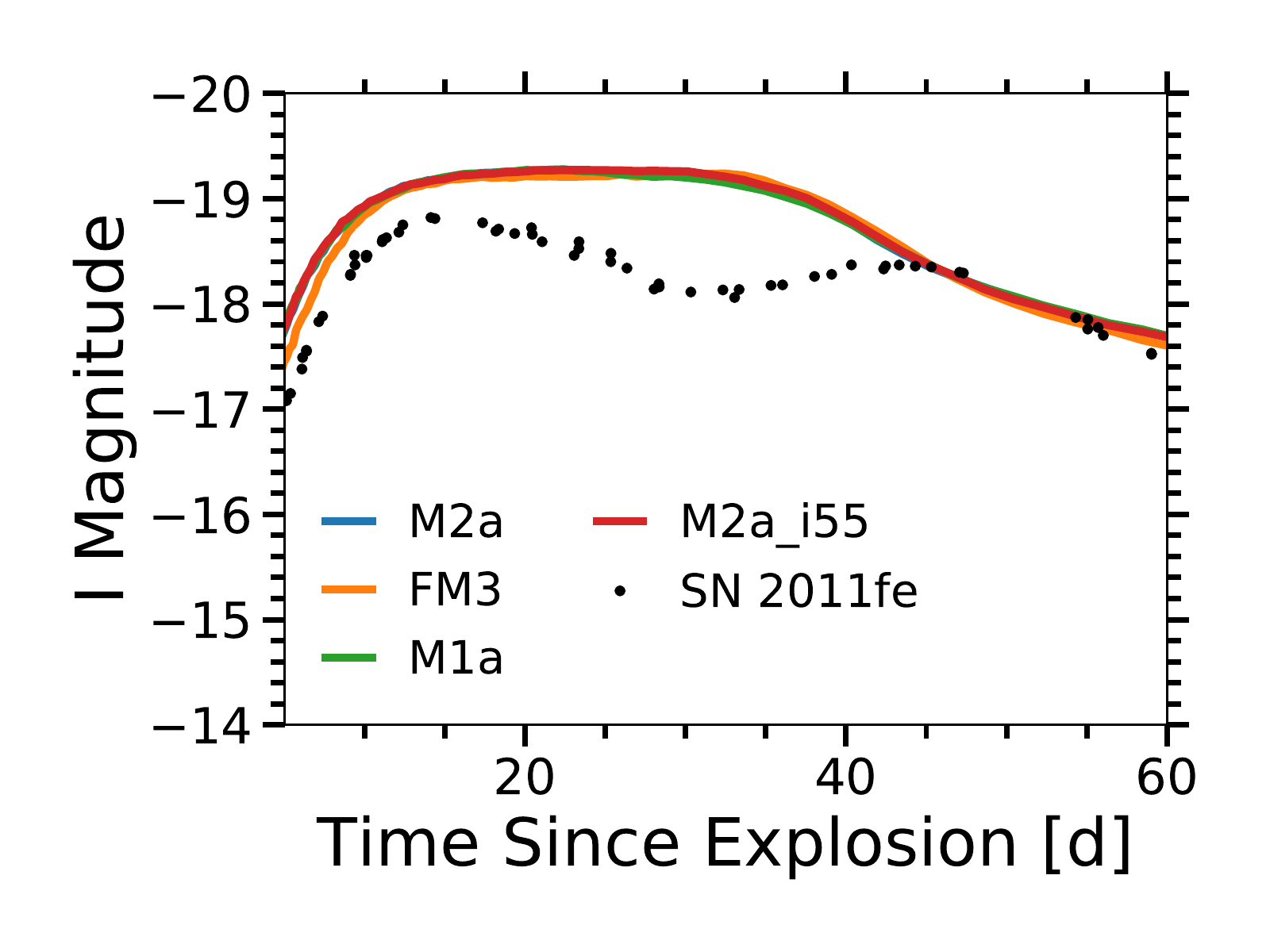}
\caption{Angle averaged U, B, V, R, and I band limited light curves for
our models as well as Model FM3 compared to the spectroscopically normal Type
Ia supernova SN~2011fe \citep{nugent2011a}.}
\label{fig:lightcurves}
\end{figure*}

The model angle averaged light curves are shown in
Fig.~\ref{fig:lightcurves}, and compared to Model FM3. The parameters for
these light curves are summarized in Table~\ref{tab:observables}.

\begin{figure*}[h]
\includegraphics[width=0.32\textwidth]{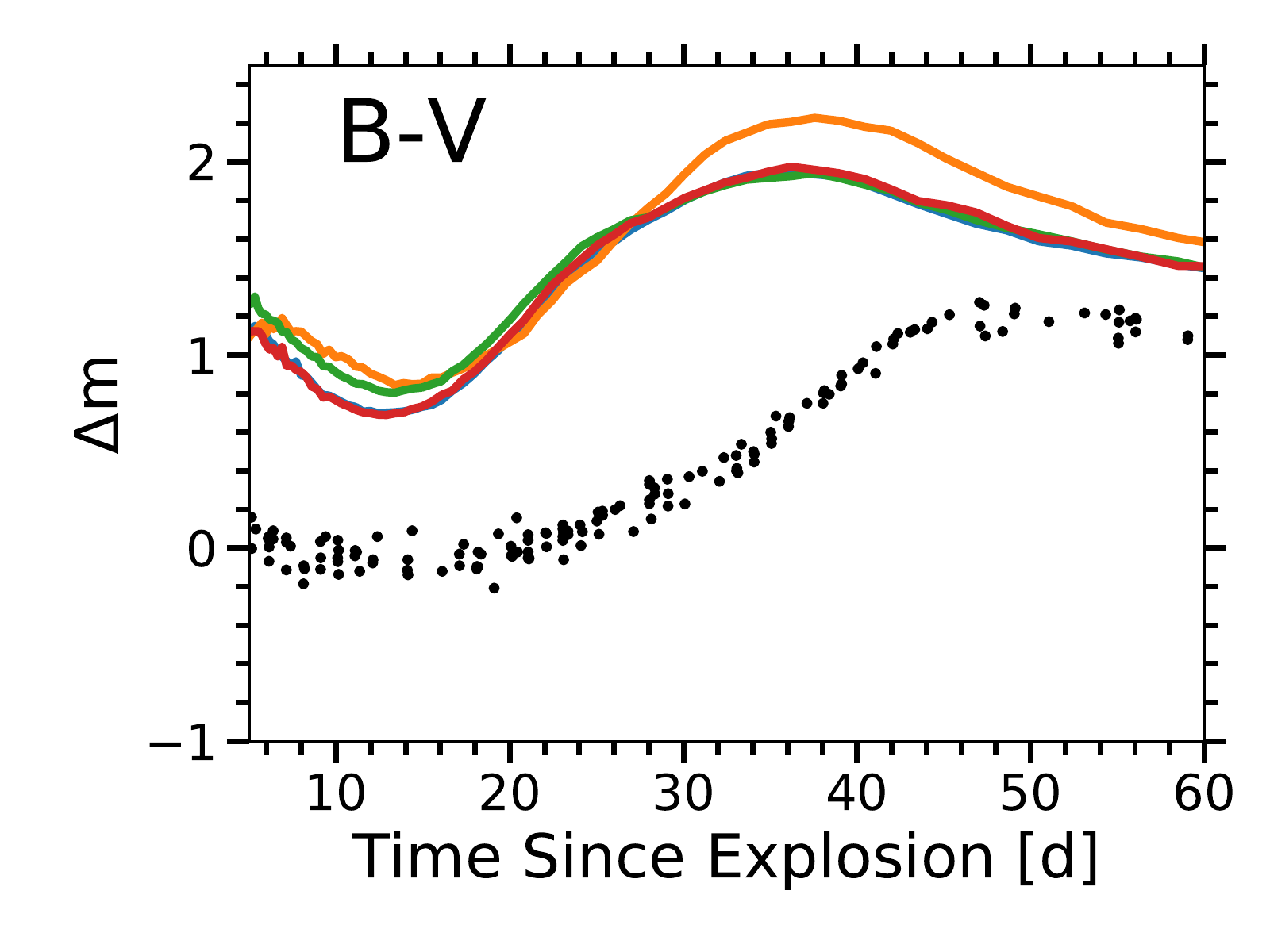}\includegraphics[width=0.32\textwidth]{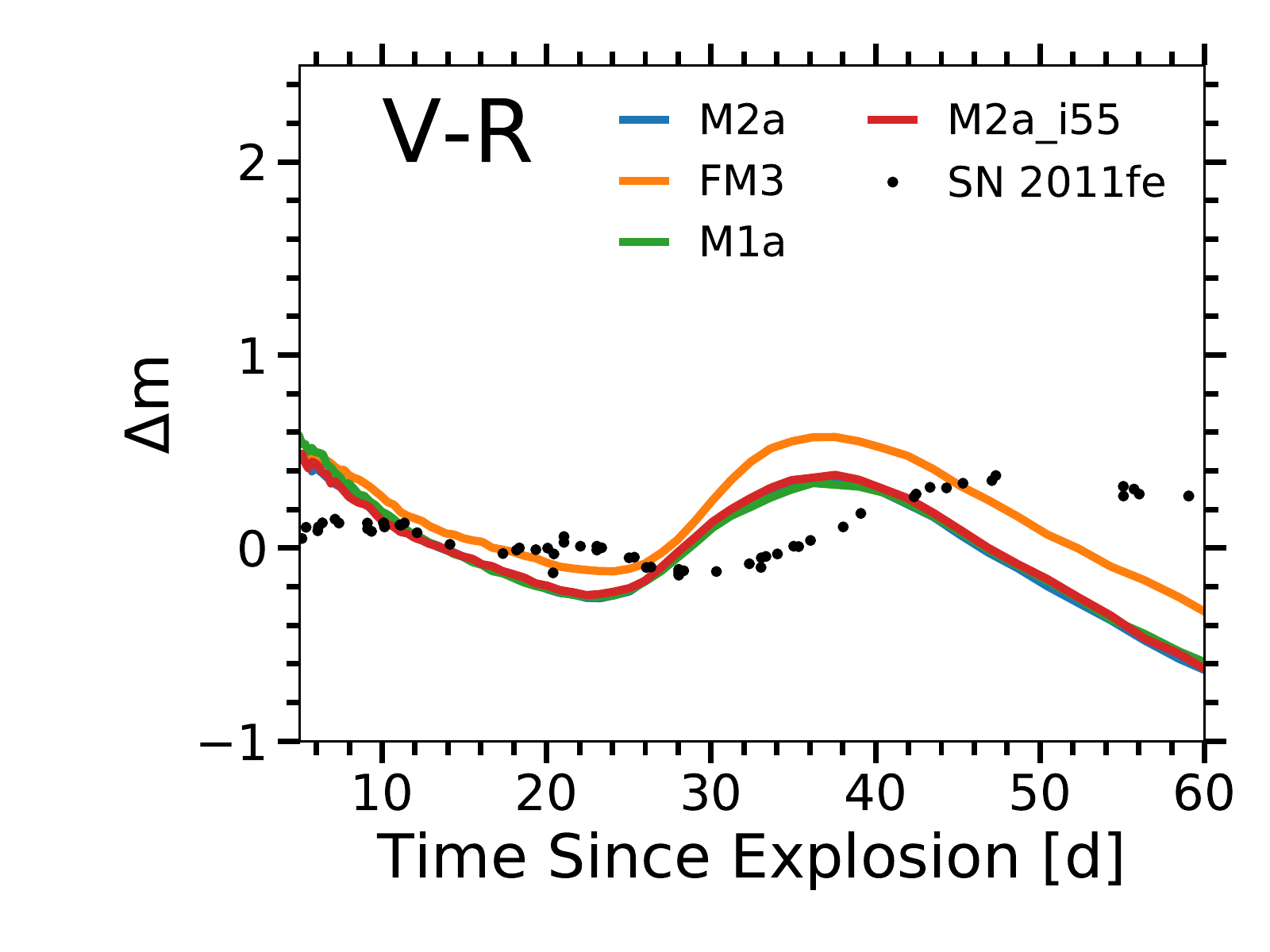}\includegraphics[width=0.32\textwidth]{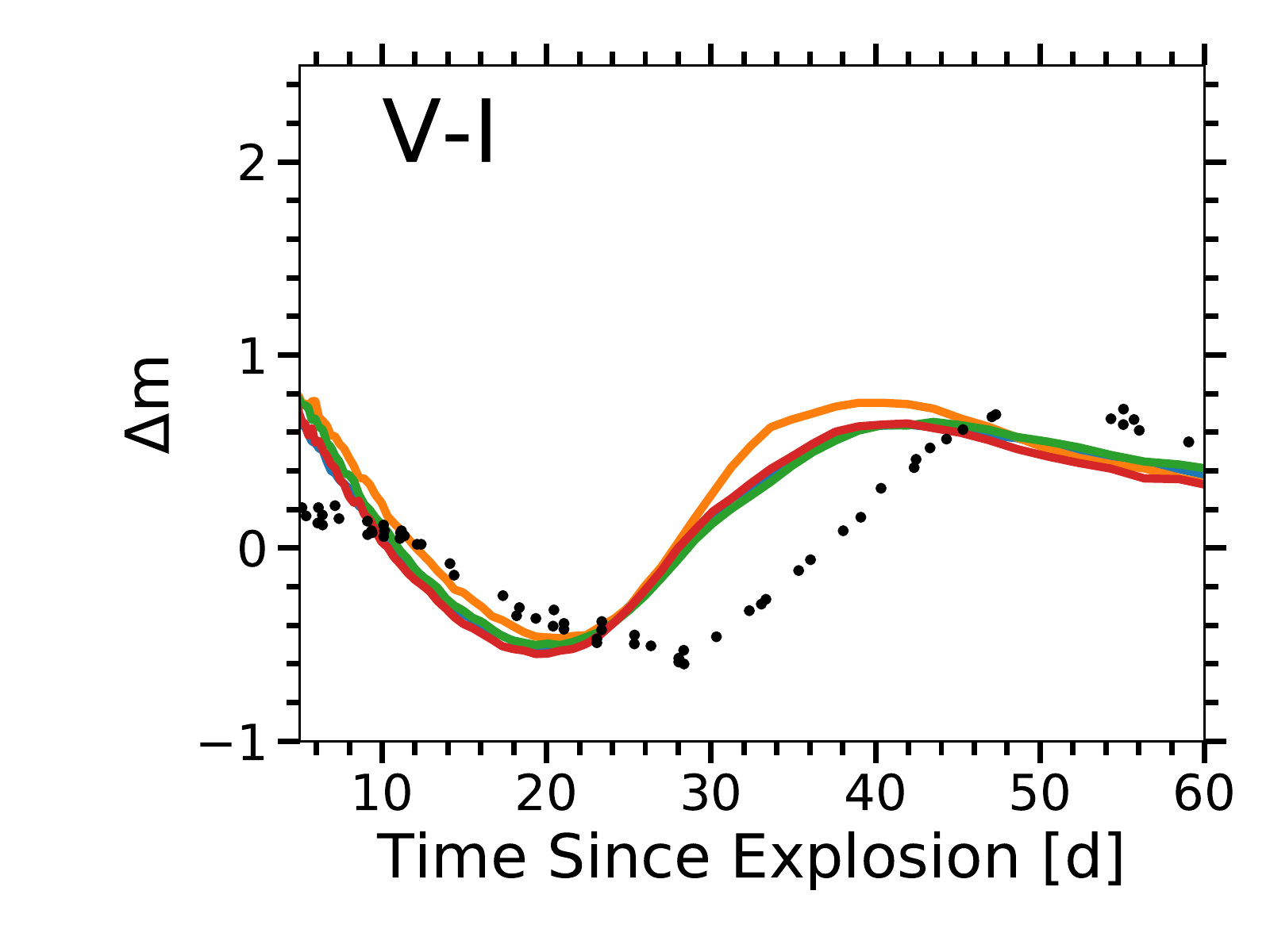}
\caption{Angle averaged B-V, V-R, and V-I color curves of the same models as in
Fig.~\ref{fig:lightcurves}. For comparison we plot the colors of the
spectroscopically normal SN~2011fe \citep{nugent2011a}.}
\label{fig:colorevolution}
\end{figure*}

\begin{table}[htbp]
    \caption{Observable parameters of Models M2a, FM3, M1a, and M2a\_i55.}
  \label{tab:observables}
  \centering
  \begin{tabular}{@{}lrrrrrr}
  \hline
  & M2a & FM3 & M1a & M2a\_i55\\ \hline
  $\Delta$m$_{15}$(B) (mag)  & 1.82 & 2.00 & 1.62& 1.83\\
  t$_{\mathrm{max}}$(B) (d)  & 16.6 & 17.7 & 16.4 & 16.6\\
  M$_{\mathrm{U,max}}$ (mag) & -18.7 & -18.4 & -18.5 & -18.7\\
  M$_{\mathrm{B,max}}$ (mag) & -18.9 & -18.6 & -18.7 & -18.9\\
  M$_{\mathrm{V,max}}$ (mag) & -19.8 & -19.7 & -19.8 & -19.8\\
  M$_{\mathrm{R,max}}$ (mag) & -19.6 & -19.6 & -19.6 & -19.6\\
  M$_{\mathrm{I,max}}$ (mag) & -19.2 & -19.2 & -19.2 & -19.2\\
  (U - B)$_{\mathrm{B,max}}$ (mag) & 0.24 & 0.24 & 0.36 & 0.26\\
  (B - V)$_{\mathrm{B,max}}$ (mag) & 0.81 & 0.95 & 0.92 & 0.82\\
  (V - R)$_{\mathrm{B,max}}$ (mag) & -0.099 & -0.016 & -0.099 & -0.093\\
  (V - I)$_{\mathrm{B,max}}$ (mag) & -0.44 & -0.39 & -0.42 & -0.47\\ \hline
\end{tabular}
\end{table}

Model M2a was chosen to be similar in mass to Model FM3, allowing a
close comparison between the outcomes of the explosion mechanisms for similar
progenitor configurations. The shapes of the light curves are similar for these
two models, despite the difference in the detonation mechanism (the scissors
mechanism for M2a and converging shock for FM3). However, in the B
band M2a is 0.3 mag brighter than FM3, peaks $\sim$ 1 day
earlier, and declines more slowly from maximum.  These effects can likely be
attributed to the higher abundance of $^{56}$Ni synthesized in the helium
detonation.

Using a 55 isotope nuclear network, as discussed in Section
\ref{sec:abundances}, rather than our standard 33 isotope network, makes only
slight differences to the light curves of Model M2a\_i55 compared to Model M2a.
Both models have the same peak brightnesses in the bands shown in
Fig.~\ref{fig:lightcurves}, and show very similar declines from maximum over 15
days in the B band. Given the similarities of these models we conclude that
using a 55 isotope network during the hydrodynamics simulation does not have a
significant effect on the model light curves.

As described in Section \ref{sec:mixing}, Model M1a investigates the influence
of core and shell mixing on the detonation ignition mechanism. In Model M1a the
abundances are reset after relaxation to bracket the effect of mixing in M2a.
This resulted in not achieving a detonation by the scissors mechanism, as for
Model M2a, but by the converging shock mechanism.  M1a peaks 0.2 days before
Model M2a, is 0.2 mag fainter at maximum in the U and B bands, and the same in
the V, R, and I bands. The differences in these explosion models make subtle
changes to the light curves, however, these differences are small, for example in
comparison to the scale of discrepancies with data (see
Section~\ref{sec:compareObservations}).

The color evolution for these models is shown in Fig.~\ref{fig:colorevolution}.
At times before maximum, the B-V colors of the models exploded by the scissors
mechanism (Model M2a and Model M2a\_i55) are marginally bluer
than those exploded by the converging shock mechanism (Model M1a and
Model FM3). At later times both the B-V colors and the V-R colors are redder
for Model FM3 than for Models M2a, M1a, and M2a\_i55.
Our models all show a very similar color evolution.  The greatest differences
seen are for Model FM3 in comparison to the new models, however, these are
still small and do not dramatically affect the level of agreement to be found
in comparison to data.

\subsection{Angle averaged spectra}
\begin{figure}[h]
\centering
\includegraphics[width=0.47\textwidth]{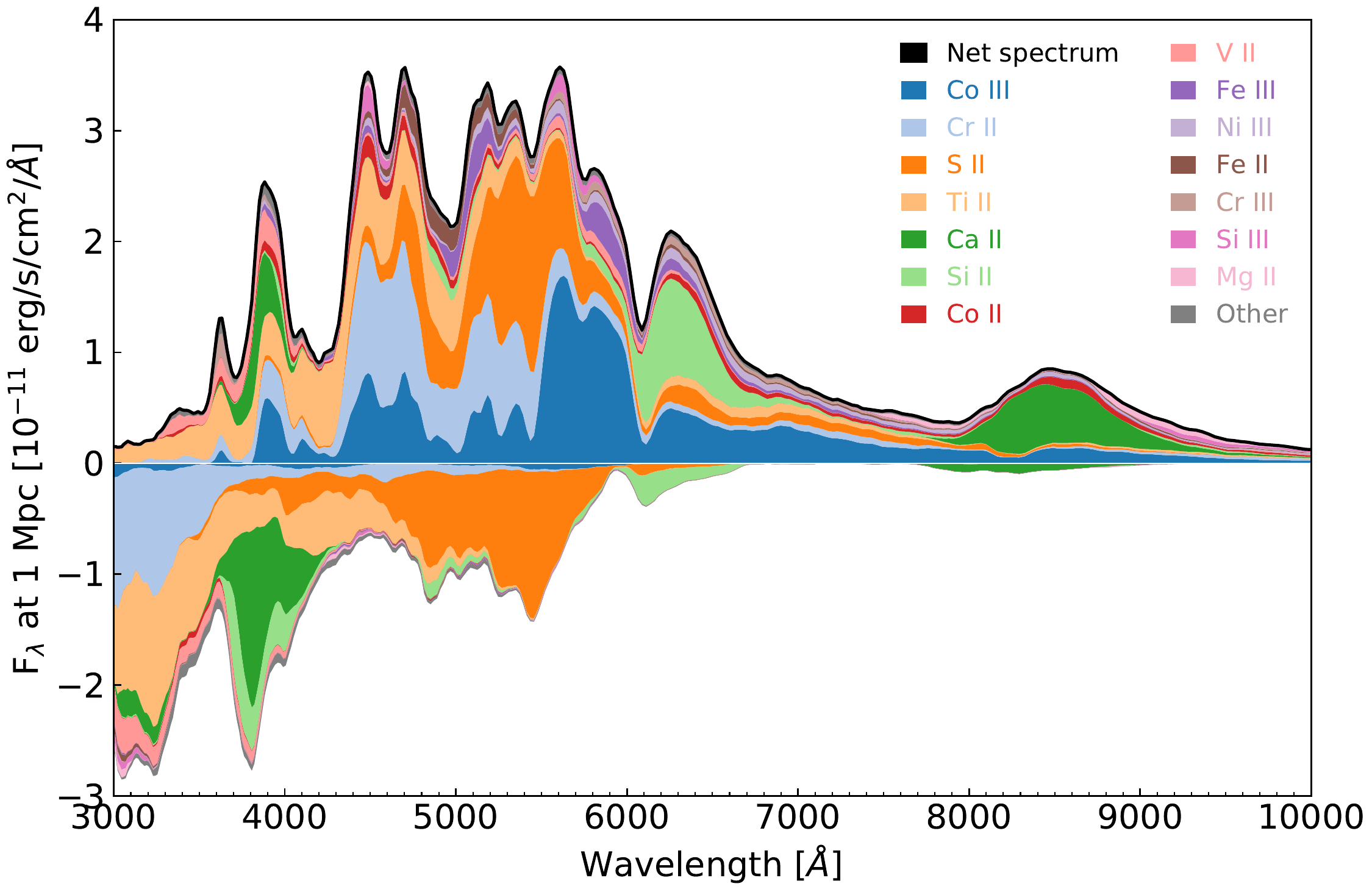}
\caption{Angle averaged emission and absorption spectrum for Model M2a
        at 18 days after explosion. The total emission spectrum is plotted in
        black. The color coding indicates the ions responsible for the emission
        and absorption.  The ions are listed in the legend of this figure in
        order of greatest contribution of flux.}
\label{fig:specemission18days}
\end{figure}

In order to understand the elemental contributions responsible for shaping the
spectra, we indicate the spectral contributions to emission and absorption,
identified by ion for Model M2a at 18~days after explosion in
Fig.~\ref{fig:specemission18days}. Specifically, in the Monte Carlo
simulations, we record details of the last interaction each escaping Monte
Carlo packet underwent. For each wavelength bin in the synthetic spectrum we
then color code the area under the spectrum in proportion to the energy
carried by packets in that bin whose last interaction was with each of the ions
considered. We also construct an equivalent histogram of color coded
contributions based on where the wavelength bin packets were prior to their last
interaction (i.e. indicating where packets last underwent
absorption/scattering/fluorescence) and plot this under the spectrum as an
indication of the key absorption processes.
This analysis confirms that the helium shell ash causes strong absorption
features in the bluer regions of the spectrum, as has been previously shown
\citep{kromer2010a}.

\begin{figure}[h]
\includegraphics[width=0.45\textwidth]{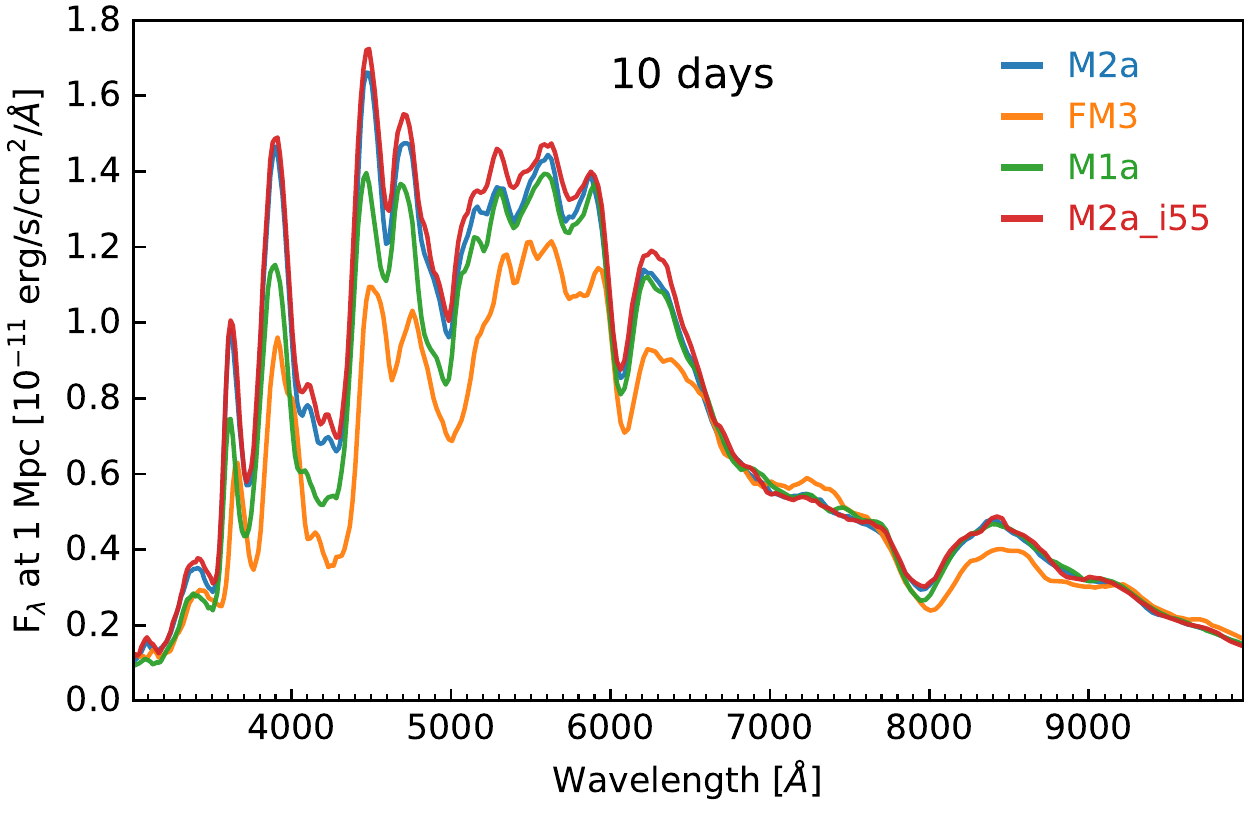}

\includegraphics[width=0.45\textwidth]{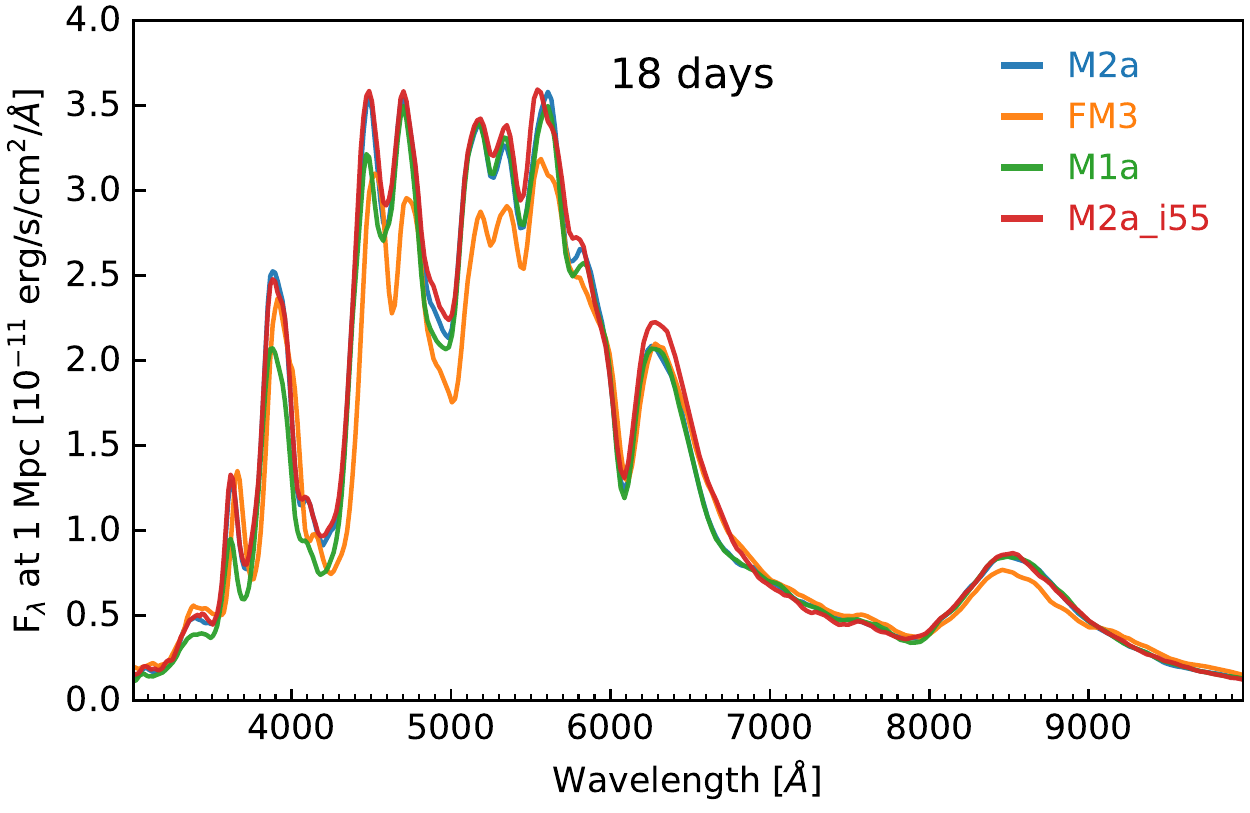}
\caption{Comparison of the spectra for Models M2a, FM3,
M1a, and M2a\_i55 at 10~days (upper panel) and 18 days (lower
panel) after maximum light.}
\label{fig:spectra_10and18days}
\end{figure}

The spectra for the models are compared in Fig.~\ref{fig:spectra_10and18days}
at 10~days and 18~days after explosion.  As expected from the similarities
between the light curves for these models, the spectra for each of the models
do not show significant differences. Again we see that the most prominent
differences are for Model FM3. At 10~days the \ion{Si}{ii} emission at
$\approx$ 6400 $\AA$ and \ion{Ca}{ii} emission at~$\approx$~8500~$\AA$ is
weaker for Model FM3, and the strength of \ion{Ti}{ii} absorption,
especially at 10 days, is strongest for Model FM3.

\subsection{Comparison objects}
\label{sec:comparisonObjects}
We compare our models to SN~2011fe \citep{nugent2011a}, SN~2016jhr
\citep{Jiang2017a}, and SN~2018byg \citep{de2019a}. Out of these, SN~2011fe is
a very well-observed Type Ia supernova of normal brightness, and is
spectroscopically normal. It is therefore a suitable benchmark to judge the
validity of our models for rather normal SNe~Ia.

SN~2016jhr was specifically suggested to have been triggered by a helium shell
detonation. It showed a prominent early optical flash $\sim 0.5$ days after
explosion, an early red and rapid color evolution, and a light curve typical of
normal brightness Type Ia supernovae, but showed strong titanium absorption,
which is typically seen in the spectra of sub-luminous supernovae.
\cite{Jiang2017a} attribute the early flash to the decays of $^{56}$Ni and
other radioactive isotopes in the outer layers of the ejecta, produced in the
helium detonation.  They ruled out interaction between the ejecta and either
circumstellar material or a companion star as the cause of the early flash due
to the early red color, as their models showed a bluer color evolution for
these scenarios.

Both the objects mentioned above (SN~2011fe and 2016jhr) are of similar peak
brightness to our models.  We also make comparisons of our models with
SN~2018byg, which -- although significantly fainter -- has also been suggested
to be the result of a helium-shell double detonation on a
sub-Chandrasekhar-mass white dwarf. The bluer regions of the spectra of
SN~2018byg show unusually strong line blanketing, with broad \ion{Ti}{ii} and
Fe-group element absorption features, and near peak they observe a deep, high
velocity ($\approx$ 25 000 km s$^{-1}$) \ion{Ca}{ii} triplet absorption
feature. The light curves of SN~2018byg are sub-luminous and similar to
SN~1991bg-like Type Ia supernovae, except for a rapid rise in r-band magnitude
within the first week from explosion. Our models are therefore systematically
too bright compared to SN~2018byg, however, we include this to make spectral
comparisons due to the similar nature of the proposed explosion scenario.  An
explosion driven by a helium detonation naturally explains the high velocity
\ion{Ca}{ii} feature, as a helium detonation produces calcium, and other
intermediate mass elements, in the outer layers of the ejecta.  \cite{de2019a}
find that the early, fast rise in the r-band light curve is consistent with the
presence of radioactive material in the outer ejecta from a helium shell
detonation.

\subsection{Comparison to observations}
\label{sec:compareObservations}

\begin{figure*}[h]
\includegraphics[width=0.32\textwidth]{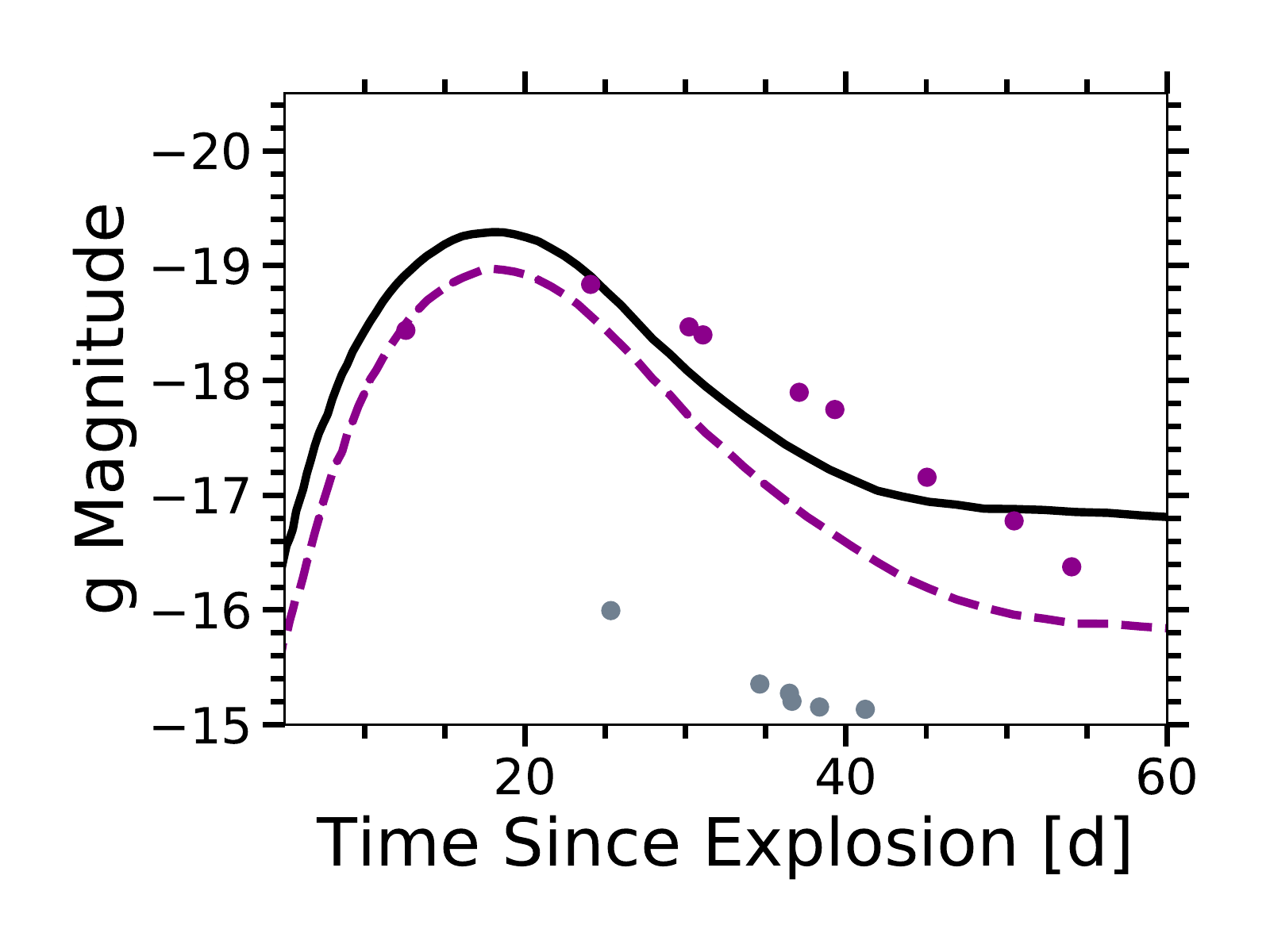}\includegraphics[width=0.32\textwidth]{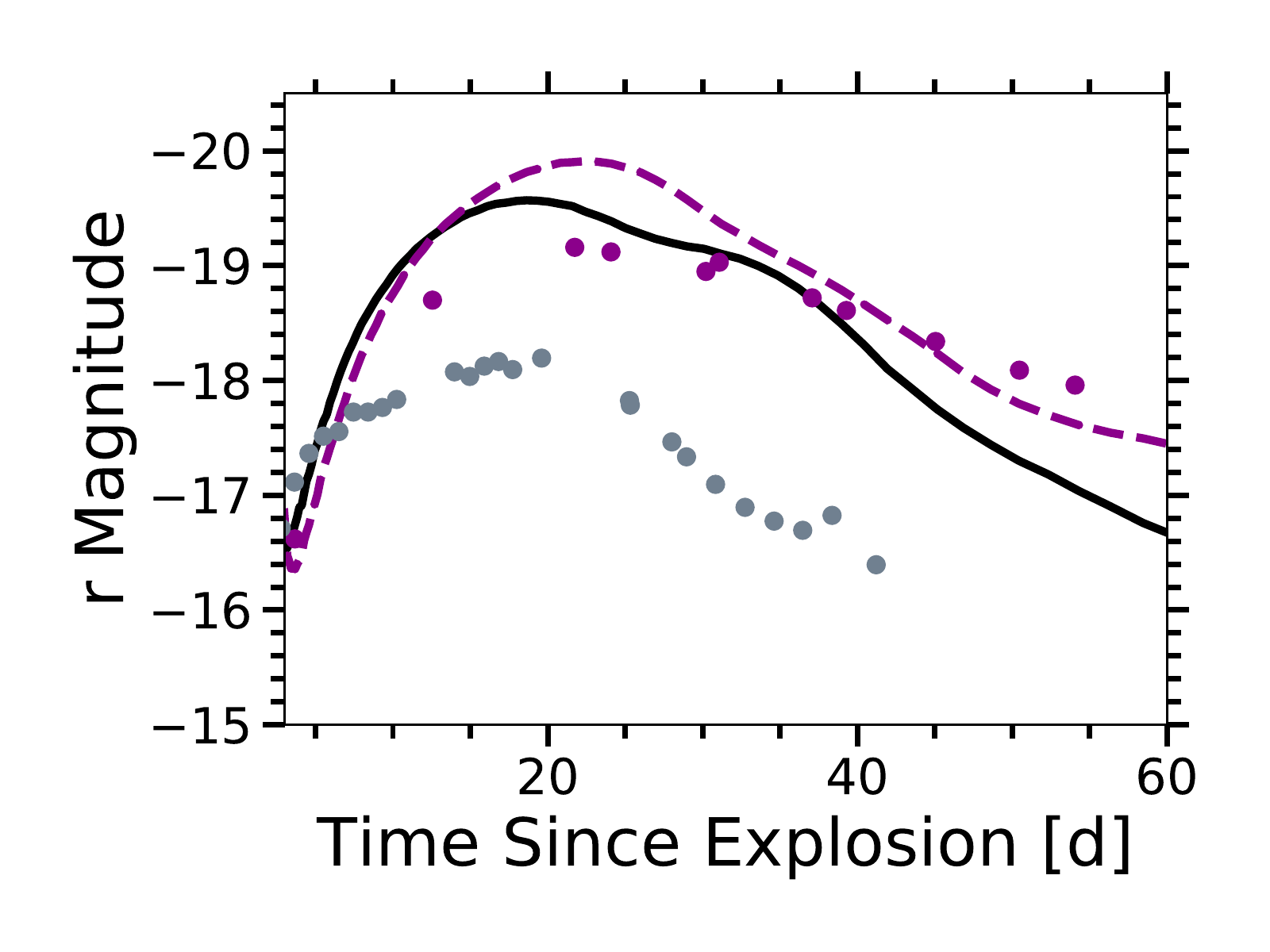}\includegraphics[width=0.32\textwidth]{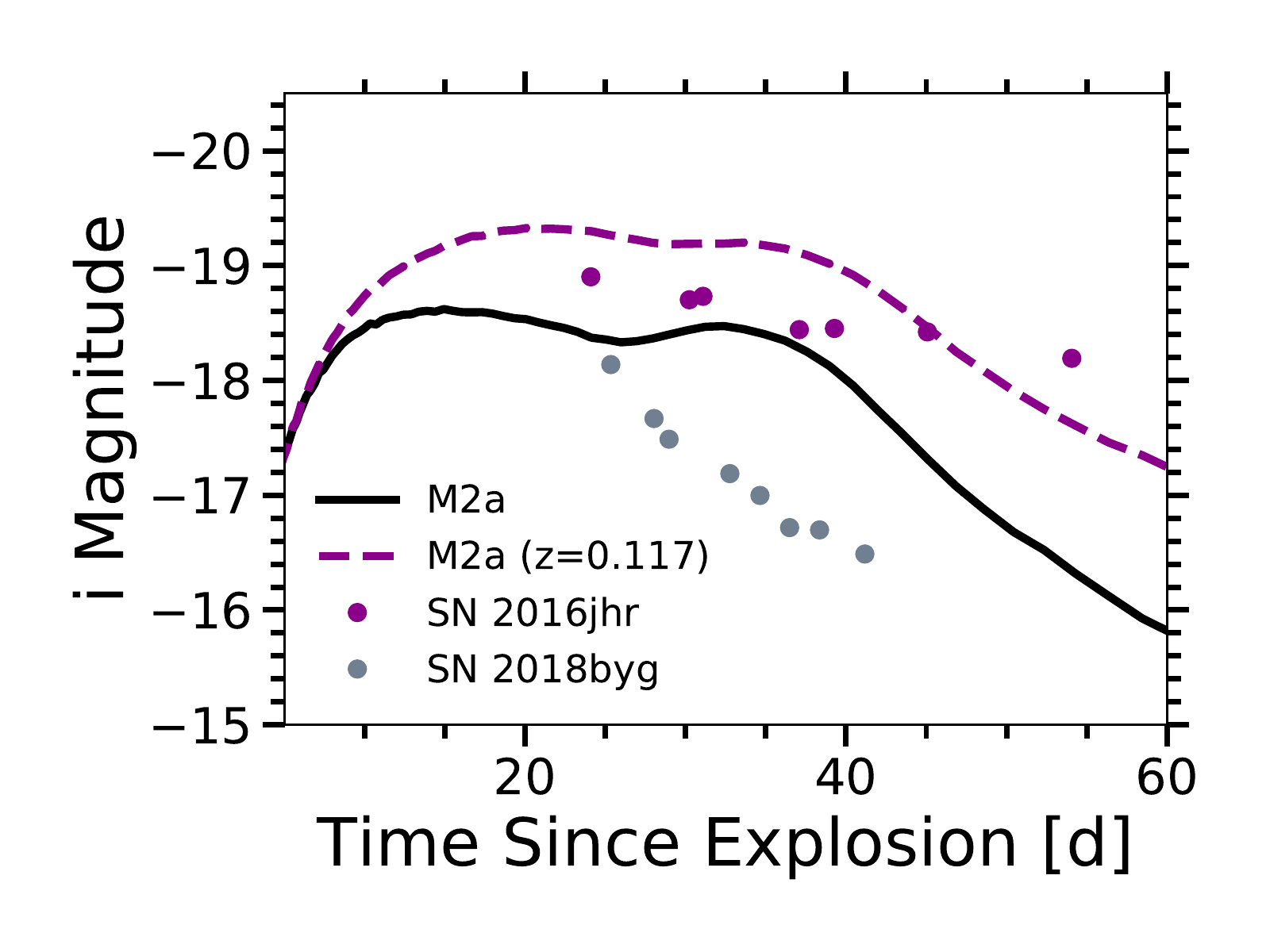}

\caption{Angle averaged g, r, and i band limited light curves of Model M2a
    compared to the light curves of SN~2016jhr \citep{Jiang2017a} and
    SN~2018byg \citep{de2019a}.  We also show the light curves of Model M2a as
    would be observed for an object at a redshift of z=0.11737 (dashed purple
    line) to make a direct comparison to SN~2016jhr.}
\label{fig:sn18byglightcurves}
\end{figure*}

\begin{figure*}[h]
\includegraphics[width=0.32\textwidth]{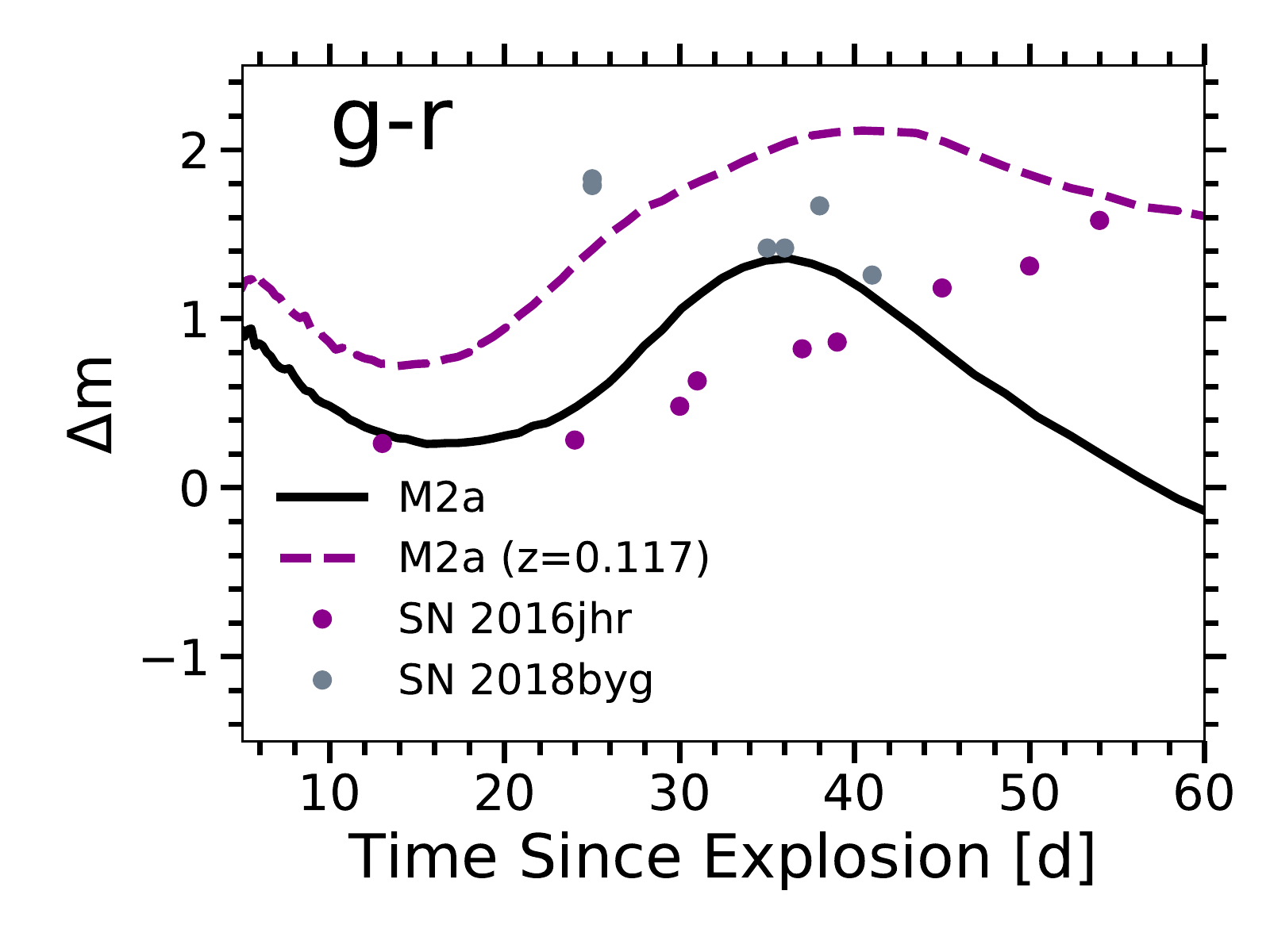}\includegraphics[width=0.32\textwidth]{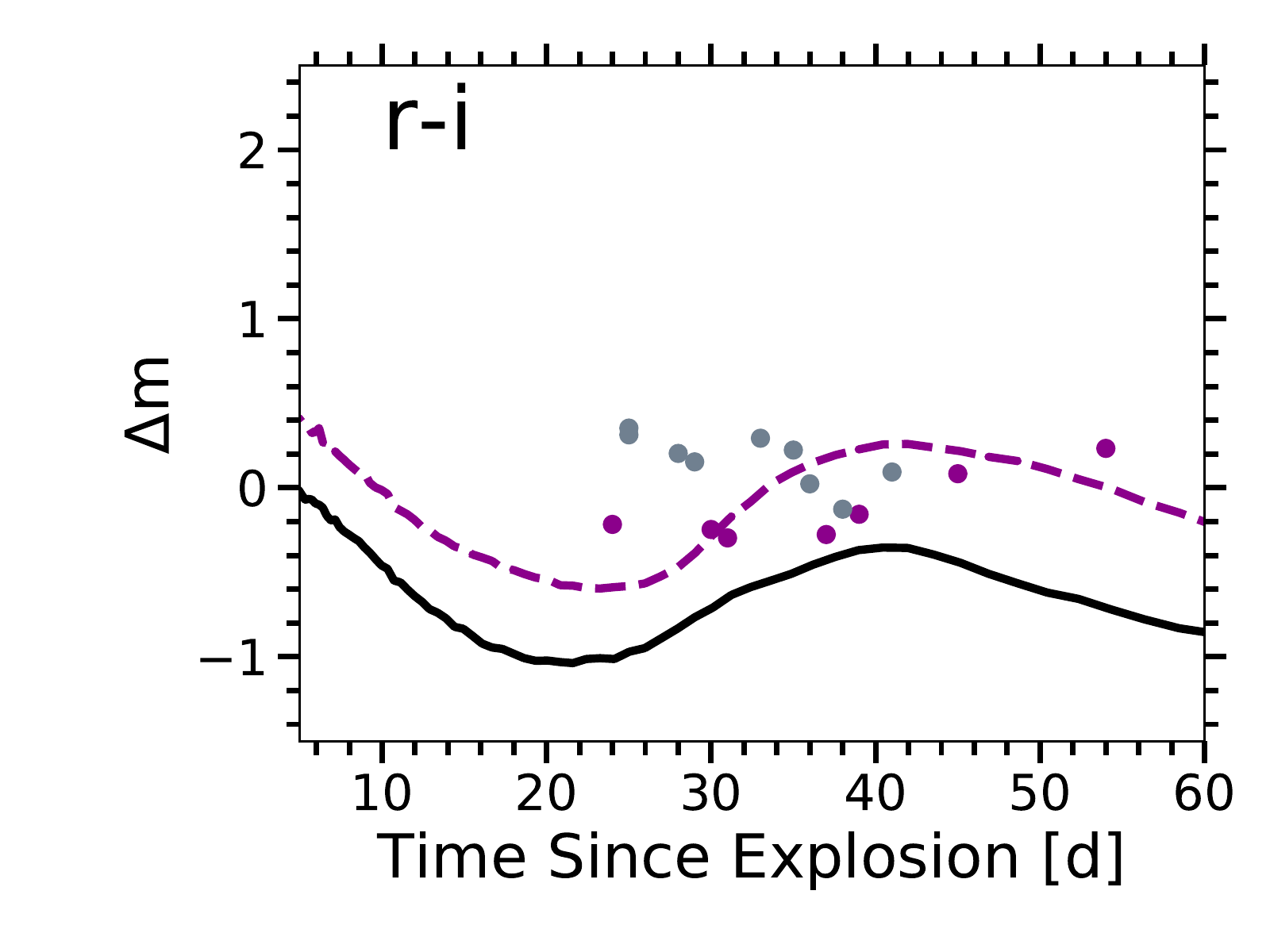}
\caption{Angle averaged g-r and r-i color curves for Model M2a compared
to the colors of SN~2016jhr \citep{Jiang2017a} and SN~2018byg \citep{de2019a}.
We also plot Model M2a at redshift z=0.11737 for comparison with SN~2016jhr.}
\label{fig:colorevolution_gri}
\end{figure*}

To compare our models with observations we first only discuss the angle
averaged light curves and spectra for our models. We discuss viewing angle
effects in Section \ref{sec:viewingangles}. We compare the model light curves
to our comparison objects, followed by a comparison of the model colors to
these objects. The spectra are discussed afterwards. We correct the spectra and
photometry for SN~2016jhr and SN~2018byg for reddening due to Galactic
extinction (given by E(B-V)$_\mathrm{MW}$ = 0.0263 mag \citep{Jiang2017a} and
AV = 0.032 mag \citep{de2019a}, respectively).  The total extinction to
SN~2011fe was found to be negligible \citep{nugent2011a}.  The spectra of all
our comparison objects are redshift corrected.

The U, B, V, R, and I band light curves of SN~2011fe are included in
Fig.~\ref{fig:lightcurves}. We find that our models are of similar brightness
in the B band until around maximum. As discussed by \cite{kromer2010a} for
Model FM3, our models decline from maximum in the B band too rapidly compared
to observed normal SNe~Ia of comparable brightness.  We also find our models to
be too faint in the U band. This region of the spectrum is strongly affected by
absorption from the helium shell ash. In the V, R, and I bands we find that our
models are brighter than SN~2011fe around peak, but are of similar brightness
at later phases.

The B~band maximum of SN~2016jhr is -18.8 mag (derived by
\citealt{Jiang2017a}), which is similar to our reference model
M2a, however, SN~2016jhr peaks at $\sim$~20 days after explosion while
Model M2a peaks at 16.6 days after explosion. Again, our
models decline more rapidly in the B band than SN~2016jhr, which has a
$\Delta$m$_{15}$(B) of $\sim$ 1 mag.

SN~2016jhr has a redshift of z = 0.11737, and SN~2018byg has a redshift of z =
0.066304. Hence the effects on the light curves due to redshift are not
negligible. To show the extent of this effect, and to make a direct comparison
to SN~2016jhr, we plot the light curves of Model M2a from our synthetic spectra
after they have been redshifted to z = 0.11737 in
Fig.~\ref{fig:sn18byglightcurves}. We also account for the time dilation at
this redshift. The g, r, and i band light curves of SN~2016jhr and SN~2018byg
are plotted, however, we note that this is not a direct comparison for
SN~2018byg given the lower redshift of this object.  Model M2a declines more
rapidly from maximum in the g~band than SN~2016jhr, and is brighter in the
r~band near maximum.  As SN~2018byg is sub-luminous, the brightnesses of our
model light curves do not match this object. Future work will include a
parameter study that investigates models of lower luminosity.

The angle-averaged time evolution of the B-V, V-R, and V-I colors are shown in
Fig.~\ref{fig:colorevolution}, and compared to SN~2011fe. The early B-V color
of the models is much too red at early phases compared to normal SNe Ia. This
was also found for the models in \cite{kromer2010a}. They argued that this was
mainly due to blanketing effects of the burning products of the helium shell.
The g-r color of Model M2a is also redder than SN~2016jhr (once corrected for
redshift, see Fig.~\ref{fig:sn18byglightcurves}).  The g-r color of SN~2018byg
around maximum is very red, (\citet{de2019a} found g-r $\approx$~2\,mag at peak
light).  This is significantly redder than is seen for the angle averaged g-r
color of Model M2a around maximum. The extreme redness near peak for this
object is a result of the strong line blanketing seen in the spectra.

As discussed in Section \ref{sec:comparisonObjects}, SN~2016jhr showed an
optical flash and red color evolution at early times, and SN~2018byg showed a
rapid early rise in the r band. \cite{Noebauer2017a} find that model FM3
produces an early peak in the U and B bands, and a pronounced shoulder in the V
and R bands within the first two days from explosion, which is consistent with
these observations. \textsc{Artis}, however, is ill suited for modeling such
optically thick conditions at these very early times. The abundances listed in
Table \ref{tab:abund} indicate that Models M2a, M1a, and
M2a\_i55 have large abundances of radioactive material present in the
outer ejecta produced by the helium shell detonation. As such, we would expect
these models to also show an early peak in the light curves at early times due
to radioactive decays. This effect should be investigated in future work.

\begin{figure}[h]
\includegraphics[width=0.45\textwidth]{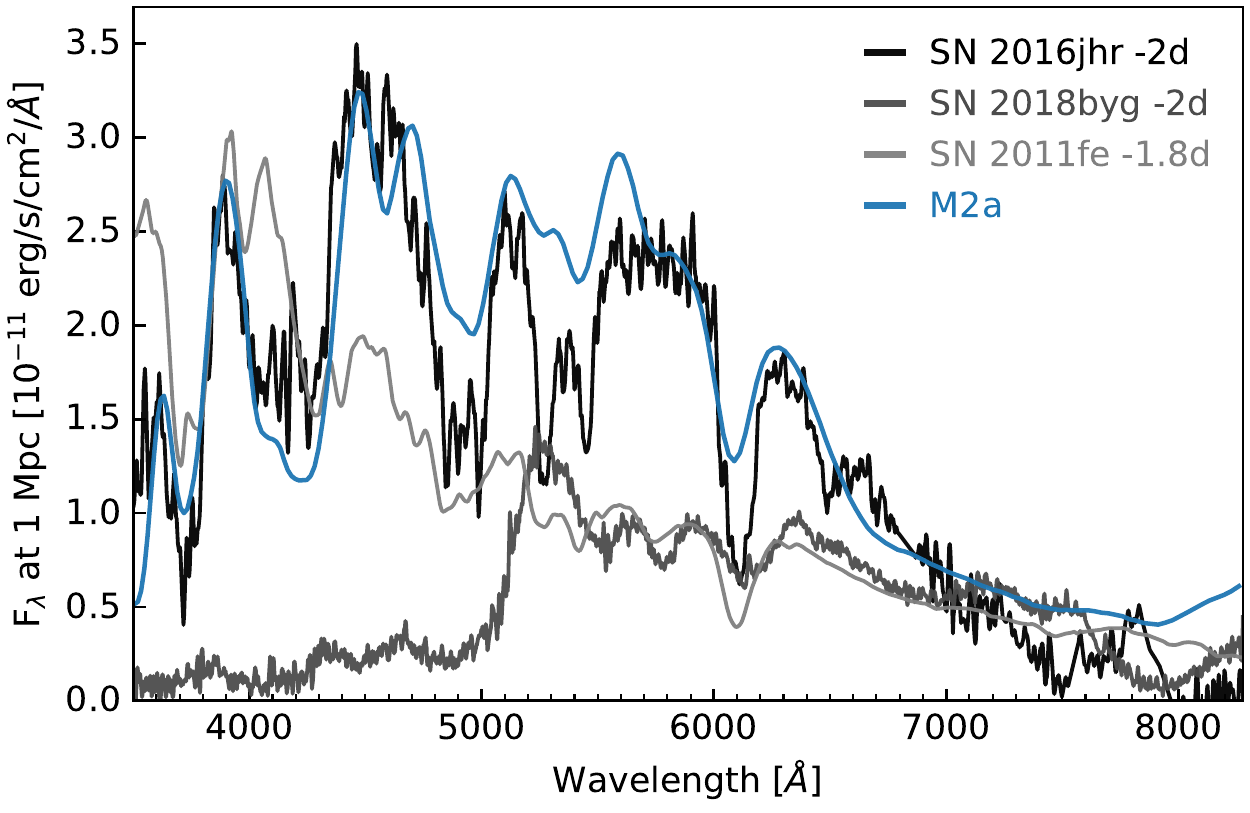}
\caption{Angle averaged spectrum of Model M2a at 15 days after
        explosion ($\sim$ 2 days before peak) compared to the spectra of
        SN~2016jhr \citep{Jiang2017a} 2 days before peak, SN~2018byg
        \citep{de2019a} 2 days before peak, and SN~2011fe \citep{nugent2011a}
        1.8 days before peak. SN~2016jhr and SN~2018byg have both been
        suggested to have been helium detonations. The spectra for SN~2016jhr
        and SN~2018byg have been de-reddened and redshift corrected, and are
        scaled to match the brightness of Model M2a.}
\label{fig:sn16jhrspectra}
\end{figure}

The angle averaged spectrum of Model M2a at 2 days before maximum is
compared to SN~2016jhr, SN~2018byg, and SN~2011fe at similar epochs, see
Fig.~\ref{fig:sn16jhrspectra}. In order to compare spectral features we scale
the spectra of SN~2016jhr and SN~2018byg to match the brightness of Model
M2a, given that the light curves in Fig.~\ref{fig:sn18byglightcurves}
show that the absolute brightnesses of these objects at this epoch are not a
close match to Model M2a. As could be anticipated from the light
curve plots and previous studies \citep{kromer2010a}, the models are not a good
match to SN~2011fe. The spectrum does, however, show some characteristic
features, such as the Si line.

Interestingly, SN~2016jhr does show similar features to Model M2a.
\cite{Jiang2017a} found prominent absorption features, such as the \ion{Ti}{ii}
trough at $\approx$ 4150 $\AA$.  All of our models show similarly strong
\ion{Ti}{ii} absorption. SN~2016jhr shows \ion{Ca}{ii} absorption around
$\approx$ 3700 $\AA$, comparable to Model M2a, and although we see
weaker \ion{Si}{ii} absorption at $\approx$ 6100 $\AA$ in Model M2a,
the velocity of this feature is similar.  While our models do show strong
absorption features in the blue, and are a reasonable match to SN~2016jhr at 2
days before peak, the angle averaged properties are not able to account for the
line blanketing seen in SN~2018byg for wavelengths blueward of
$\approx$~5100~$\AA$\,. We discuss the comparison to SN~2018byg further in
Section~\ref{sec:viewingangles}.

\subsection{Viewing angle effects}
\label{sec:viewingangles}

\begin{figure*}[h]
\centering
\includegraphics[width=0.31\textwidth]{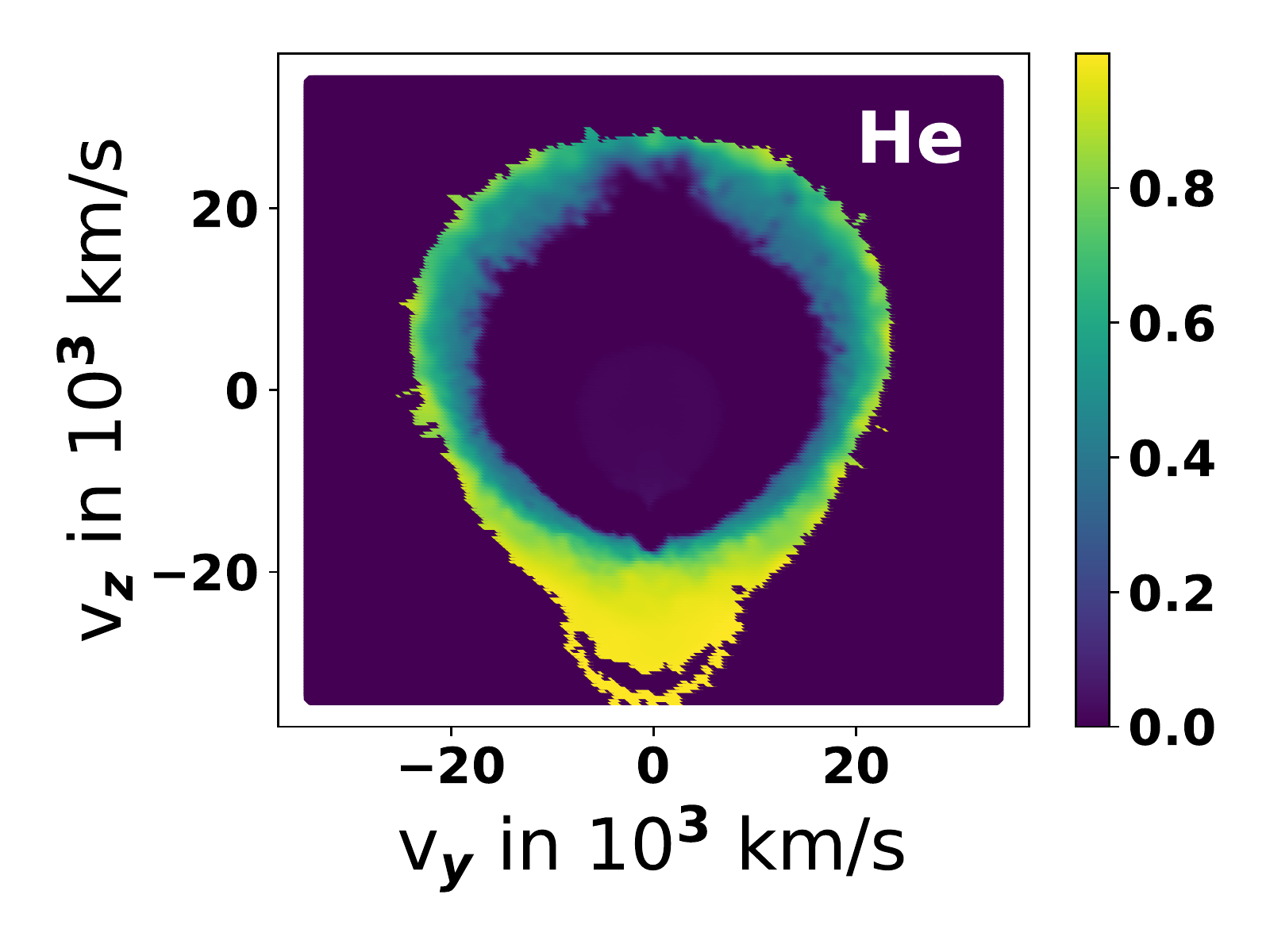}\includegraphics[width=0.31\textwidth]{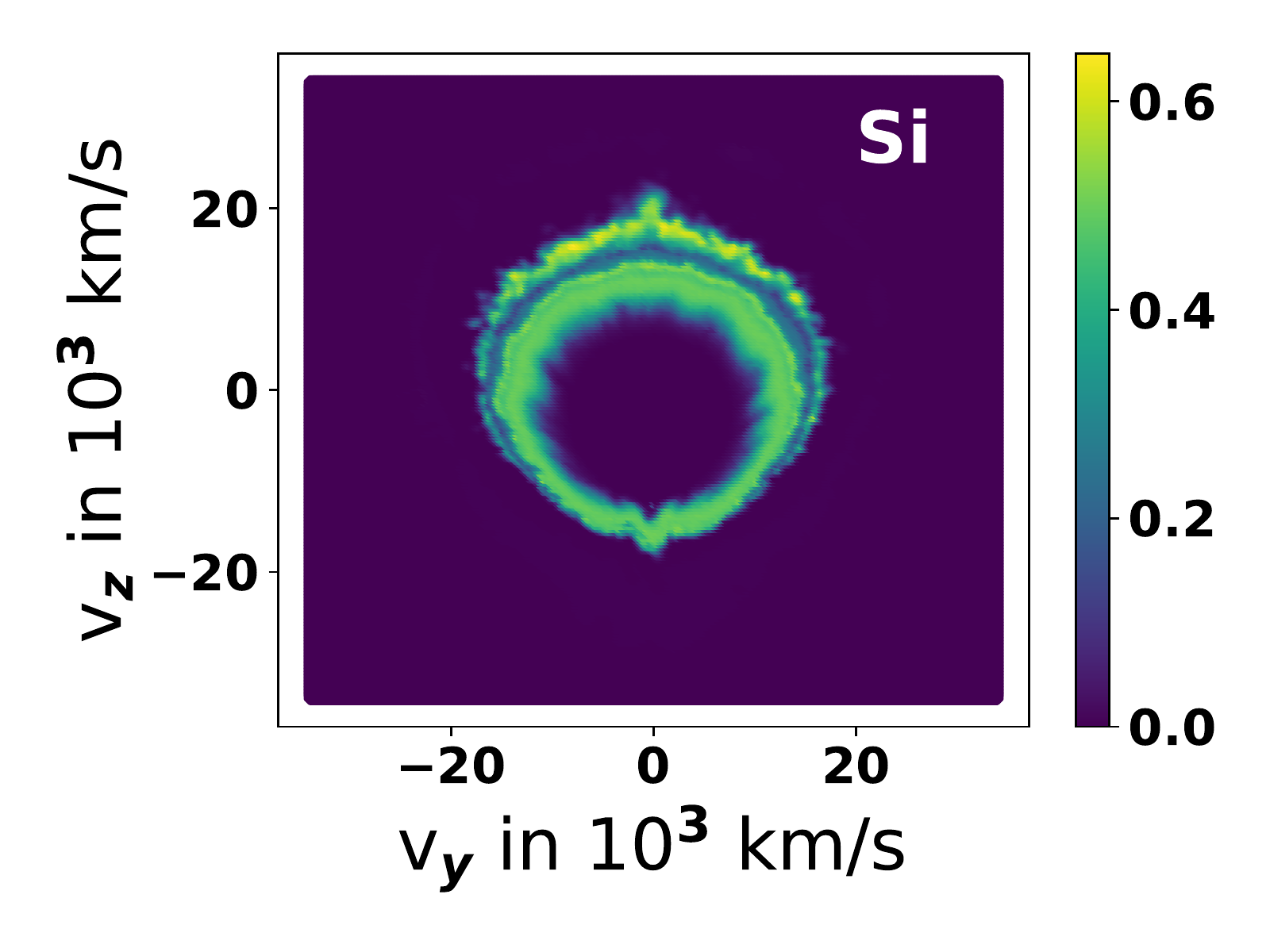}\includegraphics[width=0.31\textwidth]{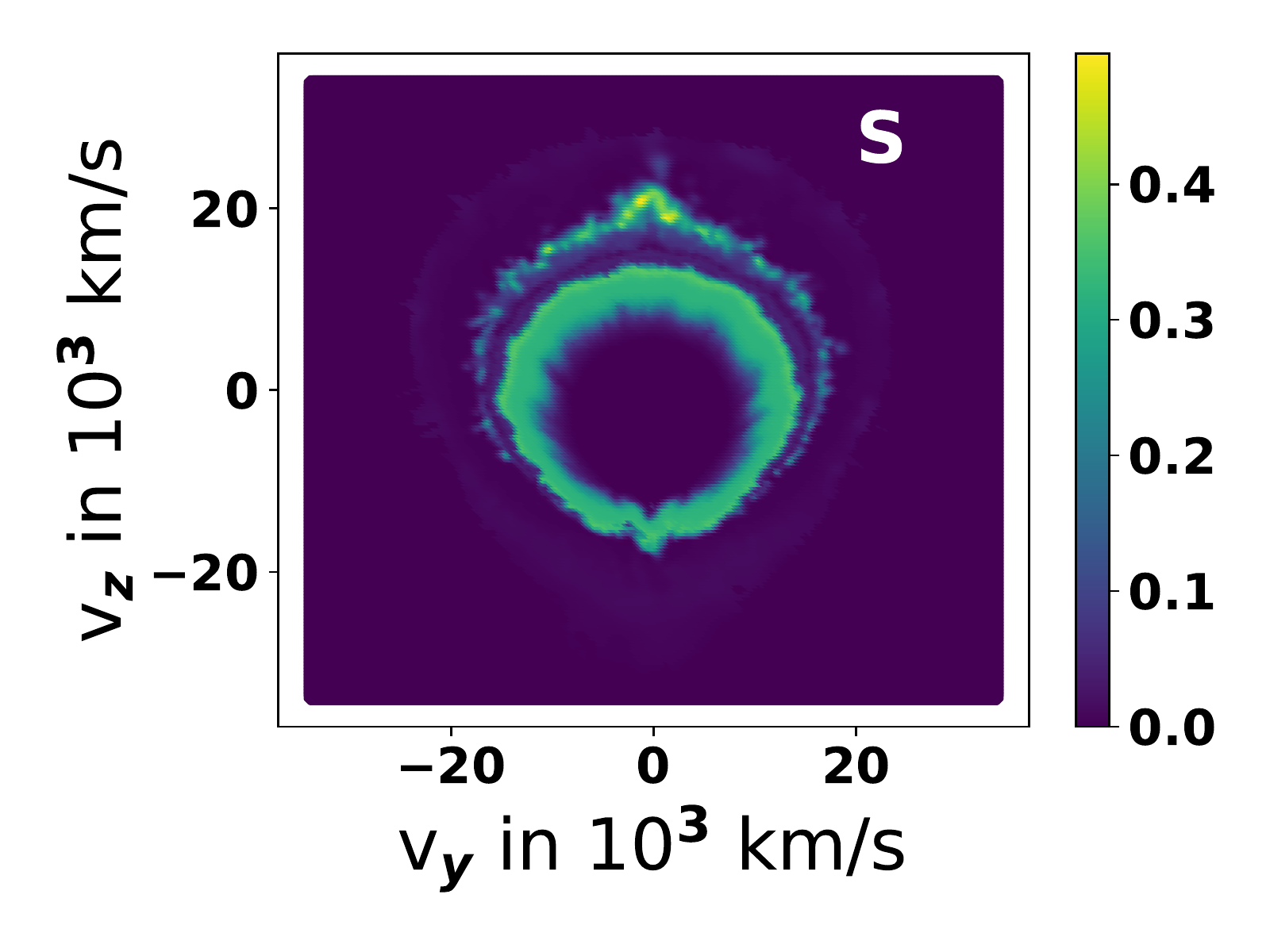}

\includegraphics[width=0.31\textwidth]{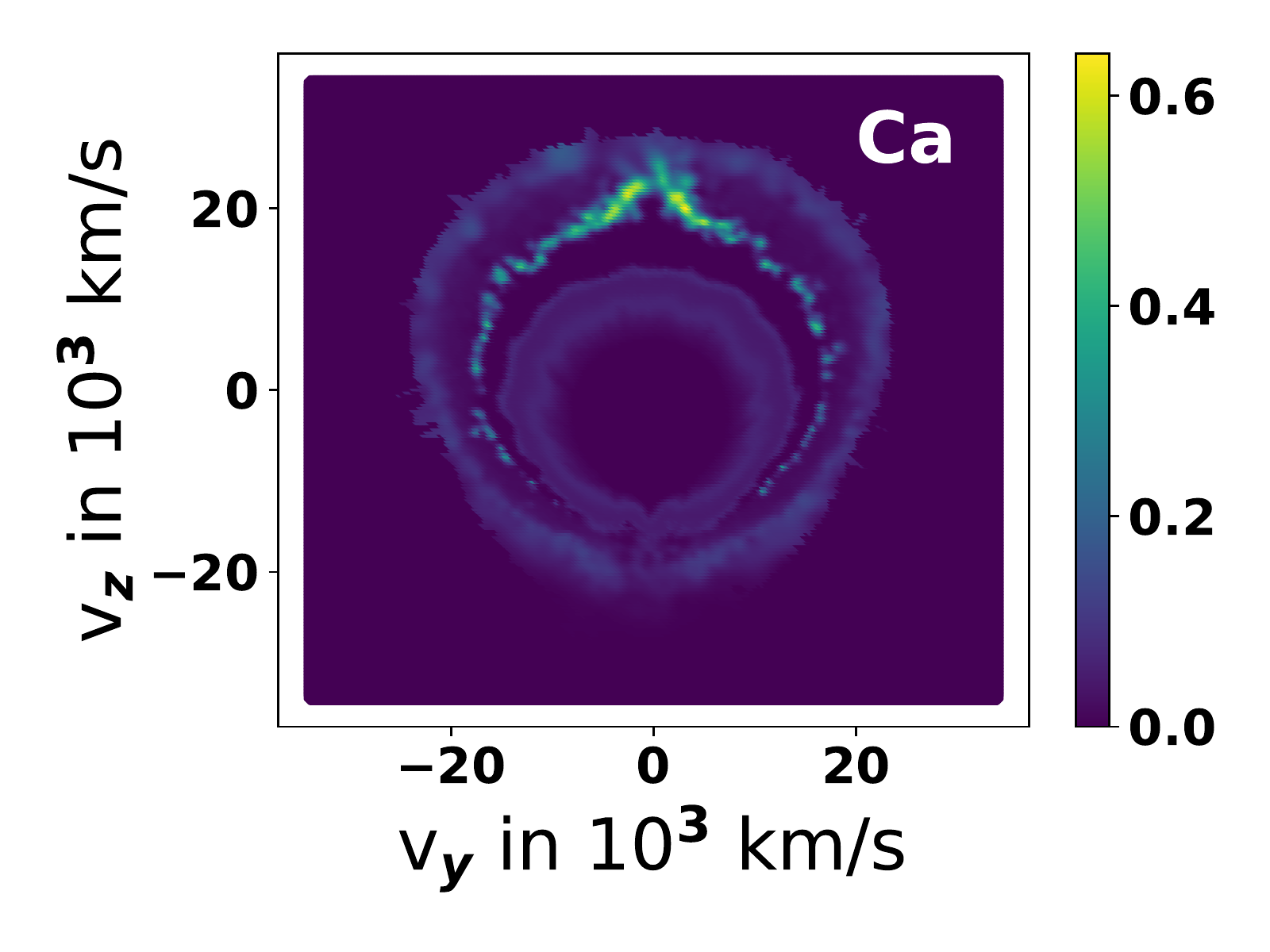}\includegraphics[width=0.31\textwidth]{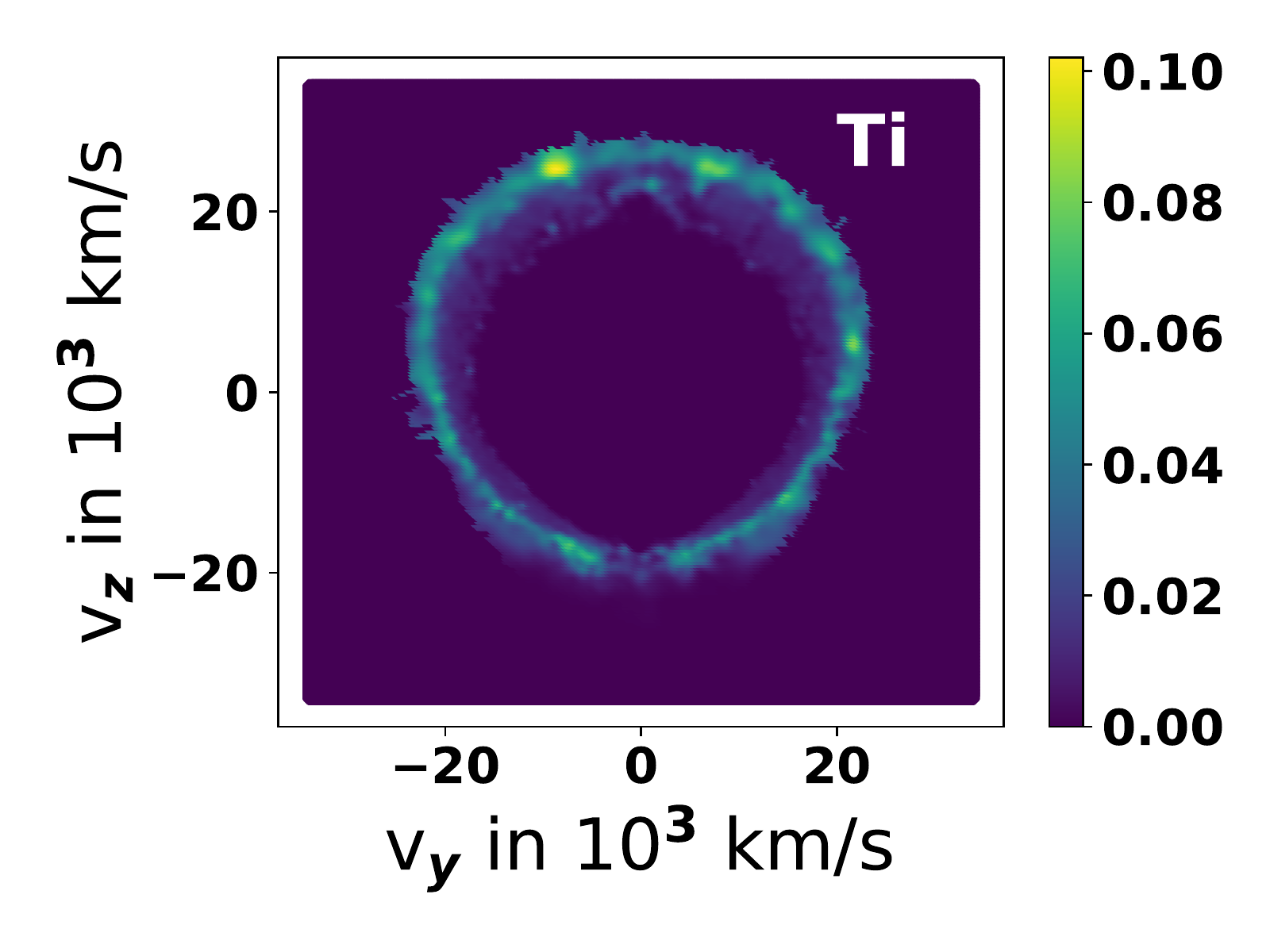}\includegraphics[width=0.31\textwidth]{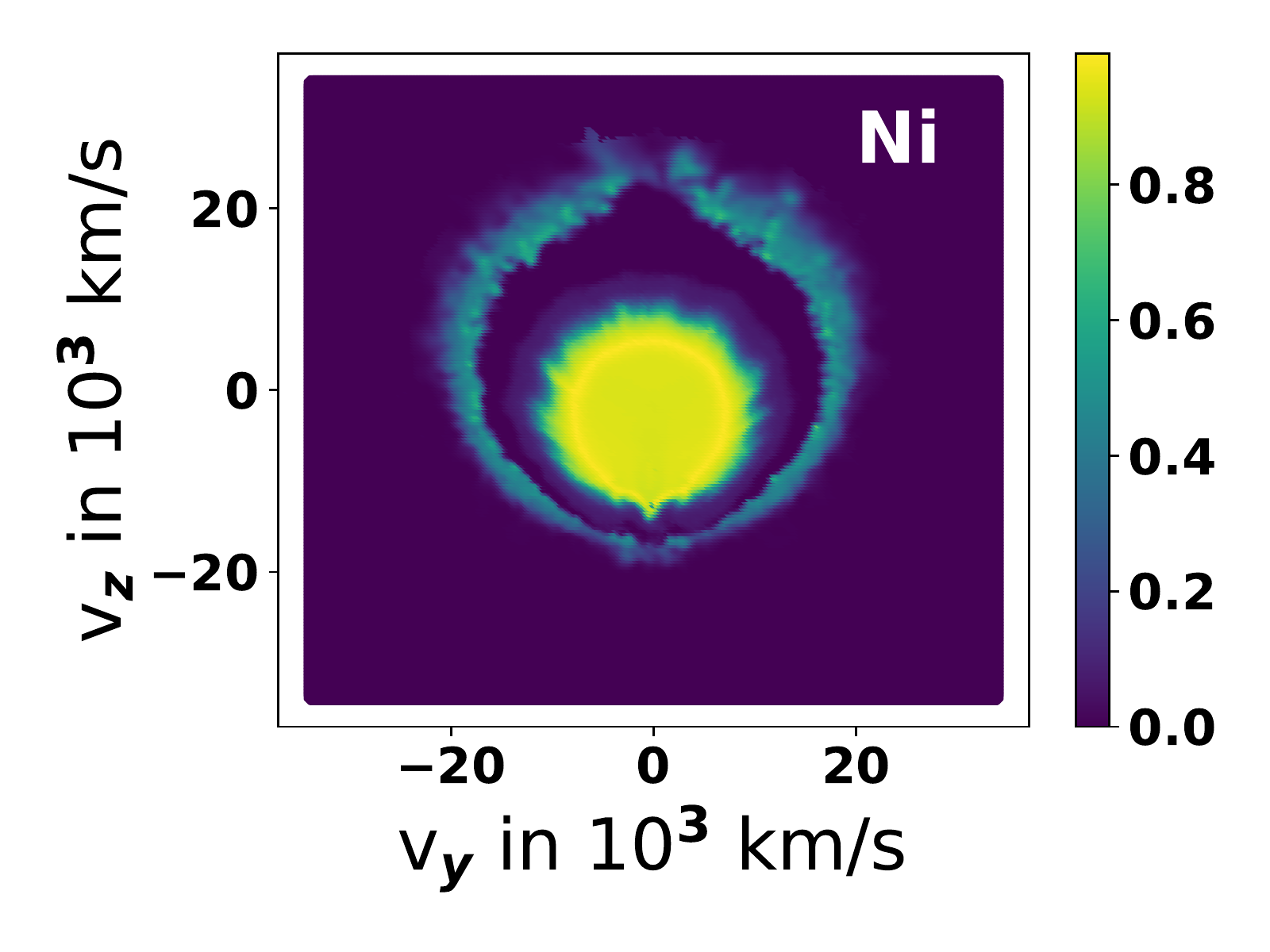}
\caption{Slice along x-axis of Model M2a showing abundances for specific
elements, where the color bar shows the mass fraction for that element.}
\label{fig:composition}
\end{figure*}

\cite{kromer2010a} show that an observer viewing the
explosion from the polar directions observes a redder spectrum, or a bluer
spectrum than viewing from equator on due to the asymmetrical distribution of
iron-group elements produced in the outer layers of their models.  It is
interesting that we see more absorption is necessary for the models to account
for the strong line blanketing of SN~2018byg, relative to the angle averaged
spectrum, as this is similar to the differences presented by \cite{kromer2010a}
between the equatorial line of sight spectrum, and the polar line of sight
spectrum where the observer is looking through an extended layer of iron-group
material produced in the helium detonation.

The off-center ignition of the scissors mechanism creates strong asymmetries in
the ejecta. Fig.~\ref{fig:composition} shows 2D slices along the $x$-axis of
Model M2a. The colors indicate the mass fraction for each of He, Si,
S, Ca, Ti, and Ni. Higher abundances of iron-group and intermediate mass
elements are synthesized around the positive $z$-axis (see Si, S, Ca in
Fig.~\ref{fig:composition}). These asymmetries create strong viewing angle
effects for this model. We find that the asymmetries in our Models
M1a and M2a\_i55 are similar to M2a. The viewing
angle effects of Model FM3 are discussed by \cite{kromer2010a}.

\begin{figure}[h]
    \begin{center}
        \includegraphics[width=0.4\textwidth]{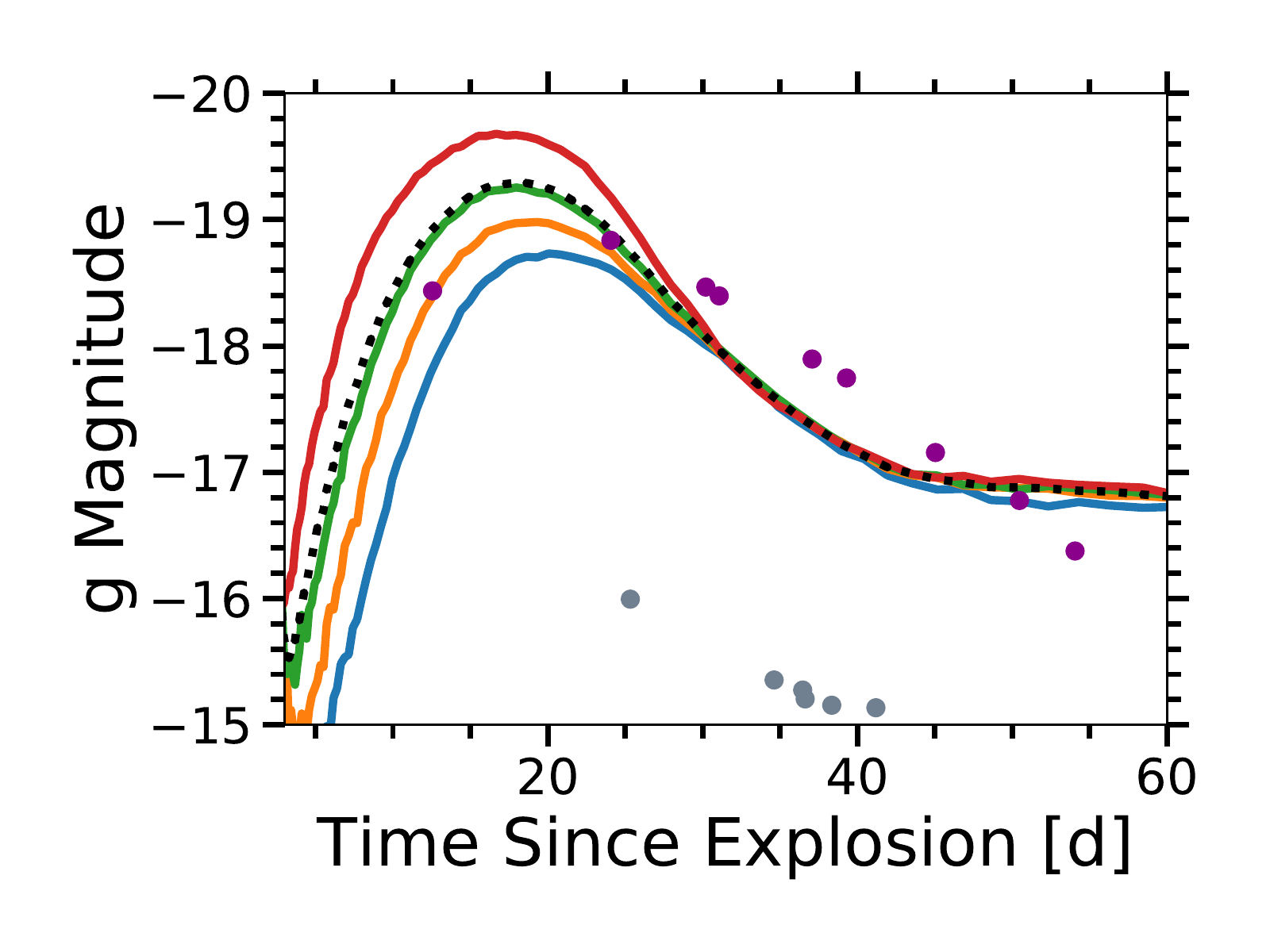}

        \includegraphics[width=0.4\textwidth]{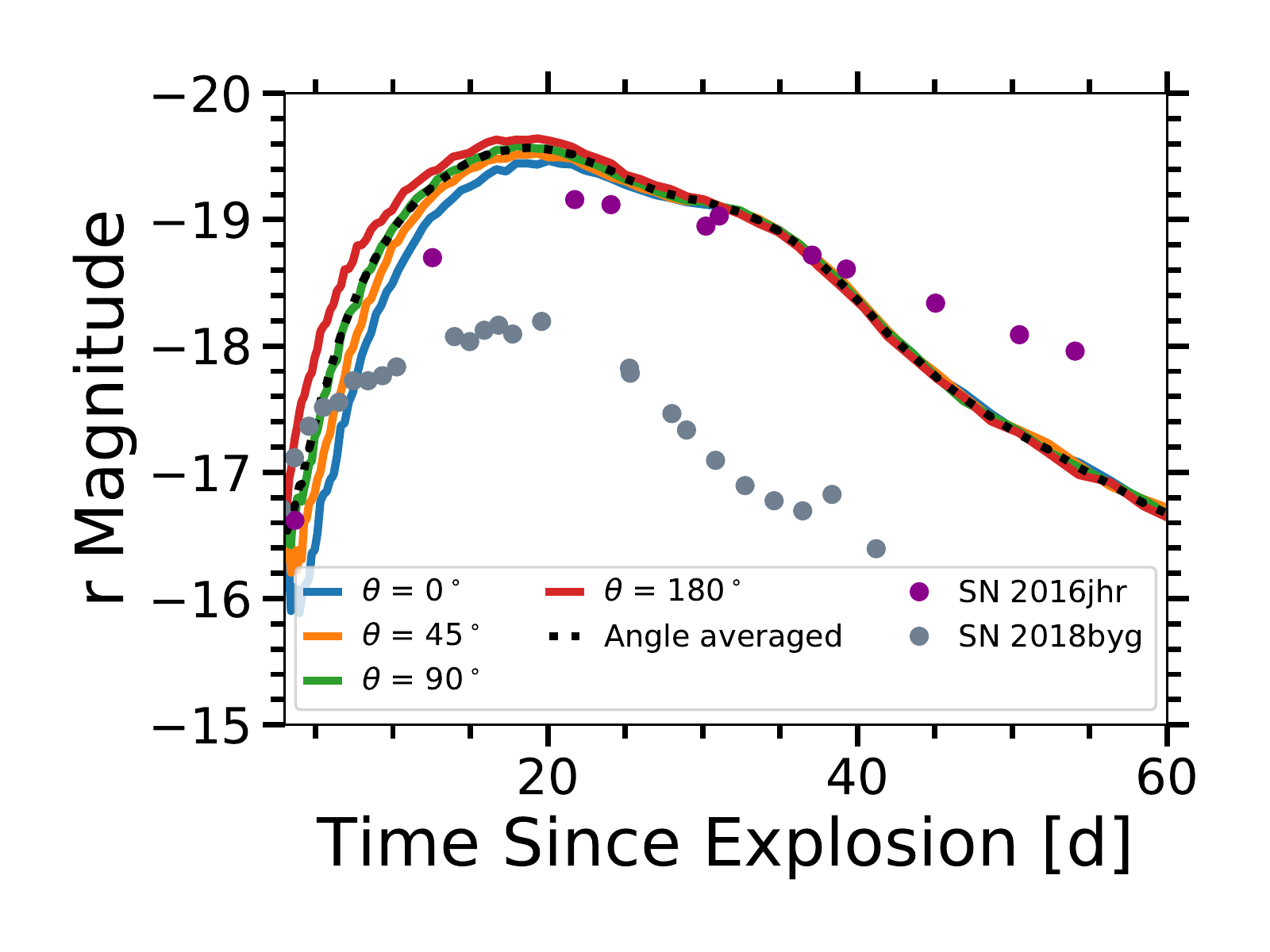}

    \end{center}

    \caption{Viewing angle dependent g and r band limited light curves for
        Model M2a, compared to the light curves for SN~2016jhr and SN~2018byg.
The angle averaged light curves are plotted (dotted black line) for reference.}
    \label{fig:angledependentlightcurves}
\end{figure}

\begin{figure}[h]
\begin{center}
\includegraphics[width=0.45\textwidth]{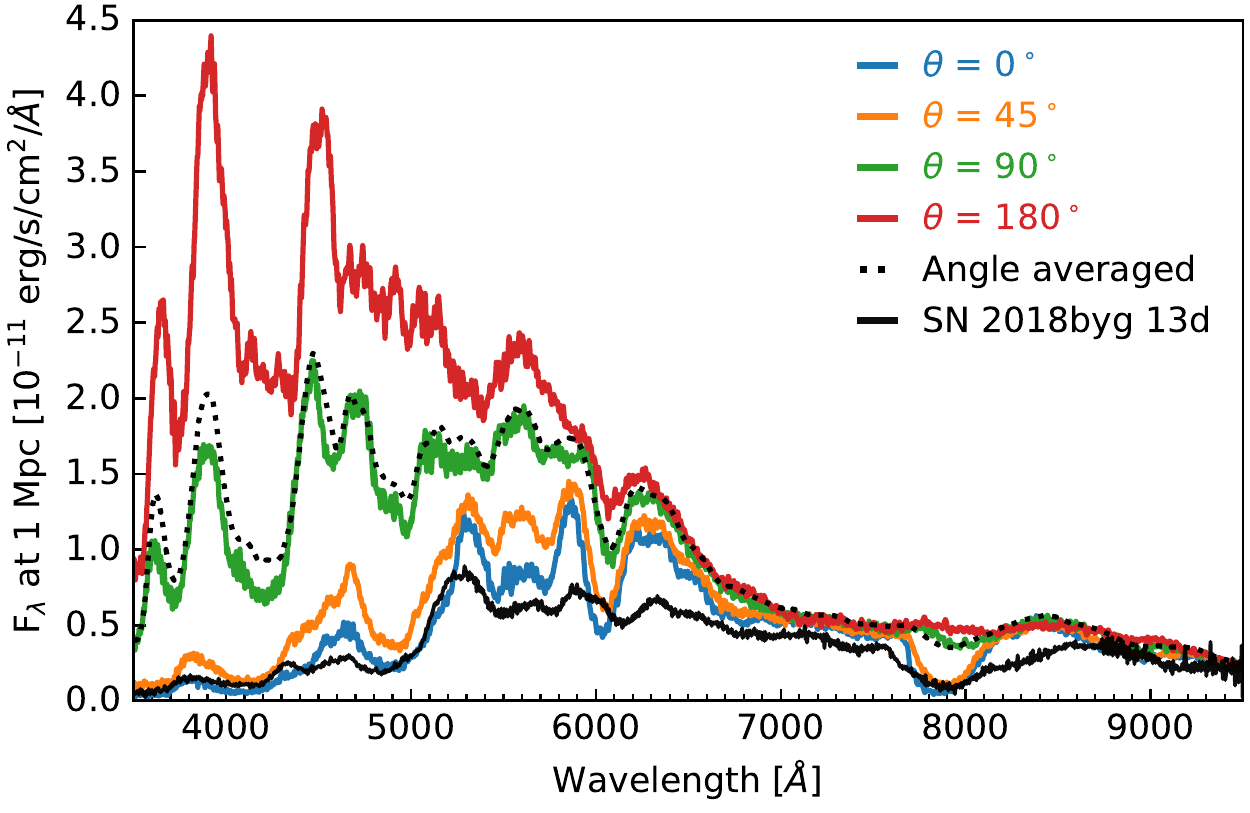}
\end{center}

\caption{Viewing angle dependent spectra for Model M2a at 12 days after
        explosion.  We show the spectra for viewing angles of $\theta$ =
        0$^\circ$, 45$^\circ$, 90$^\circ$, and 180$^\circ$, and plot the angle
        averaged spectrum for comparison.  Also plotted is the spectrum of
SN~2018byg at a similar epoch (13 days after explosion). The spectrum of
SN~2018byg has been de-reddened and redshift corrected. }
\label{fig:angledependentspectra}
\end{figure}

We show in Fig.~\ref{fig:angledependentlightcurves} the angle dependent light
curves in the g and r bands for model M2a, compared to the light
curves of SN~2016jhr and SN~2018byg. The g band light curves are more strongly
affected by the asymmetries in the ejecta, as we see a difference of $\sim$ 1
mag in peak brightness between the angles presented here. It is apparent that
the level of absorption viewed by an observer is a strong function of
orientation. As discussed in Section \ref{sec:compareObservations} the degree
of absorption in the angle averaged spectrum of our models is not sufficient to
account for the level of line blanketing observed for SN~2018byg. However, the
high degree of blanketing required by SN~2018byg is reproduced for the most
extreme lines of sight, plotted in Fig.~\ref{fig:angledependentspectra}, which
shows the viewing angle dependent spectra calculated for Model M2a at
12 days after explosion, and the spectrum of SN~2018byg at a similar epoch (13
days). The higher abundance of heavy elements (see Fig.~\ref{fig:composition})
in the direction of an observer looking toward the helium ignition point on
the positive z-axis ($\theta$ = 0$^\circ$) causes stronger absorption than for
the angle averaged spectra, as is seen in Fig.~\ref{fig:angledependentspectra}.
The spectrum viewed at 45$^\circ$ is similar to $\theta$ = 0$^\circ$. These
spectra also show strong line blanketing in the blue regions of the spectra,
and deep \ion{Ca}{II} absorption features. At $\theta$~=~180$^\circ$
significantly less absorption is seen due to the lower abundance of heavy
elements. The spectrum seen at $\theta$~=~90$^\circ$ is similar to the angle
averaged spectrum.  These results highlight the necessity of multi-dimensional
simulations.

\section{Conclusions}
\label{sec:summary}
In this work we describe a detonation ignition mechanism for the double detonation
scenario that has previously not received much attention: the scissors mechanism.
Most simulations carried out so far are in 1D or 2D
\citep[e.g.,][]{woosley1994a,bildsten2007a} with only \citet{moll2013a}
performing 3D simulations. Here 3D simulations were carried out using the
\textsc{Arepo} code. Its adaptive mesh allows us to study the evolution of the
helium shell detonation with high spatial resolution.

A detonation ignition mechanism is detected: The detonation and shock waves
propagate in the helium shell and in the core. At the point when the detonation
wave in the helium shell converges opposite to its ignition spot, high enough
densities and temperatures are reached in a large enough volume to ignite a
second detonation. This detonation propagates into the core and leads to its
complete incineration. This core detonation ignition mechanism differs from the
converging shock mechanism of \citet{fink2007a,fink2010a} and
\citet{moll2013a}. In our models, we assumed a non-rotating progenitor
configuration. Rotation introduces a symmetry axis and \cite{garcia2018b} find
that a He detonation ignition far from this axis blurs the convergence of the
detonation wave on the antipode. They conclude, however, that an ignition of
core detonation is still likely to occur. We note that if the scissors
mechanism discussed here fails, the mechanism associated with the converging
shock detonation would occur.

We find that the scissors mechanism is independent of the WD mass. However, the
profile of the transition between CO core and He shell is important (see
Sec.~\ref{sec:mixing}). A change in the mixing of carbon into the He
shell can result in a different shell composition and detonation ignition
mechanism.

The simulations show that the mechanism is robust for different resolutions and
that the energy release is converged. However, as the carbon
detonation cannot be fully resolved the carbon ignition is partly a numerical
effect. Nevertheless, as critical values found in previous work
\citep{roepke2007a,seitenzahl2009b} are reached in more than one cell it is
reasonable to say that the detonation is physical.

The final abundances from the helium detonation of Model M2a show
differences of about one order of magnitude for many isotopes compared to FM3.
The difference in the $^{56}$Ni abundance in our Model M2a is a result
of the different physical conditions in the setup and burning treatment.
\citet{roepke2017a} describes the details of the combustion processes. It
should be noted that \citet{fink2010a} consider a different ignition mechanism
for the second detonation and their simulations are in 2D. The final total
$^{56}$Ni abundance, however, is of the same order of magnitude and corresponds
to a mass in the expected range for a SN Ia
\citep{stritzinger2006b,scalzo2014a}.

Radiative transfer calculations were carried out using the radiative transfer
code \textsc{Artis}. We present the synthetic observables of Models
M2a, M1a, and M2a\_i55, and we compare these to Model
FM3 and to observed SNe Ia. We find that despite the differences in the
explosion models their light curves and spectra show no significant
differences. While these models show some differences compared to Model FM3,
these are small considering the apparent discrepancies with observations.

As was found by \cite{kromer2010a}, our models are too red compared to
observations of normal SNe Ia. This is particularly obvious in the B-V color,
where our models are redder than the spectroscopically normal SN~2011fe at all
epochs considered. Additionally, our models are only able to match some
spectral features, such as the Si II line, typical for normal SNe Ia near
maximum light.

We also compare our models to SN~2016jhr \citep{Jiang2017a} and SN~2018byg
\citep{de2019a}, which are unusual SNe~Ia specifically suggested to have been
triggered by helium shell detonations. We find that the g-r color of our
models is redder than SN~2016jhr near maximum light, however, we find that the
extreme redness of SN~2018byg around maximum is significantly redder than our
models due to the line blanketing observed for SN~2018byg.

We find that the near maximum spectrum of SN~2016jhr is a reasonable match to
the angle averaged spectrum of Model M2a, which is similar to the equatorial
line of sight. In particular the strong absorption due to intermediate mass
elements seen at the blue wavelengths of the spectrum is similar to that found
for Model M2a. The angle averaged spectrum for Model M2a does not produce
enough line blanketing to account for that seen in SN~2018byg, however, the
most extreme lines of sight are able to reproduce this level of absorption at
the blue end of the spectrum.
\\ \noindent

Here we have proven that the amount of core-shell mixing is an important
parameter which influences the details of the carbon detonation ignition.  We
find that the double detonation scenario includes a further carbon detonation
ignition mechanism --~namely the scissors mechanism~-- which did not receive
much attention before and strongly depends on the amount of mixing. However, we
have only investigated a limited set of parameters.  Therefore it is necessary
in the future to conduct a parameter study. The dependence of the final yields
and stability of the detonation ignition mechanism on the mass of the helium
shell and CO core, as well as metallicity and carbon abundance in the shell
will be studied.

\begin{acknowledgements} We thank our referee for insightful comments that
    helped to greatly improve this article. This work was supported by the
    Deutsche Forschungsgemeinschaft (DFG, German Research Foundation) --
    Project-ID $138713538$ -- SFB 881 (``The Milky Way System'', subproject A10).
    SG, STO, RP, MK, and FKR acknowledge support by the Klaus Tschira
    Foundation.  SAS acknowledges support form the UK STFC through grant
    ST/P000312/1.  IRS was supported by the Australian Research Council through
    Grant FT160100028.  NumPy and SciPy \citep{oliphant2007a}, IPython
    \citep{perez2007a}, and Matplotlib \citep{hunter2007a} were used for data
    processing and plotting.  The authors gratefully acknowledge the Gauss
    Centre for Supercomputing e.V.  (www.gauss-centre.eu) for funding this
    project by providing computing time on the GCS Supercomputer JUWELS
    \citep{juwels2019} at J\"{u}lich Supercomputing Centre (JSC).

This work was performed using the DiRAC Data Intensive service at Leicester,
operated by the University of Leicester IT Services, which forms part of the
STFC DiRAC HPC Facility (www.dirac.ac.uk). The equipment was funded by BEIS
capital funding via STFC capital grants ST/K000373/1 and ST/R002363/1 and STFC
DiRAC Operations grant ST/R001014/1.  This work also used the Cambridge Service
for Data Driven Discovery (CSD3), part of which is operated by the University
of Cambridge Research Computing on behalf of the STFC DiRAC HPC Facility
(www.dirac.ac.uk). The DiRAC component of CSD3 was funded by BEIS capital
funding via STFC capital grants ST/P002307/1 and ST/R002452/1 and STFC
operations grant ST/R00689X/1. DiRAC is part of the National e-Infrastructure.
This work was supported by computational resources provided by the Australian
Government through the National Computational Infrastructure (NCI) under the
National Computational Merit Allocation Scheme.

\end{acknowledgements}

\bibliographystyle{aa}

\end{document}